\documentclass[12pt]{article}

\title{Multivariate Subordination using Generalised Gamma Convolutions with Applications to Variance Gamma Processes and Option Pricing\footnote{This
research was partially supported by ARC grants DP1092502,
DP0988483 and DP160104037.}}

\author{Boris Buchmann\thanks{Research School of Finance, Actuarial Studies \& Statistics,
Australian National University,
ACT 0200,
Australia,
Phone (61-2) 6125 7296,
Fax (61-2) 6125 0087,
email: Boris.Buchmann@anu.edu.au
}
\and
Benjamin Kaehler\thanks{John Curtin School of Medical Research, Australian National University,
ACT 0200,
Australia, email: Benjamin.Kaehler@anu.edu.au}
\and
Ross Maller\thanks{Research School of Finance, Actuarial Studies \& Statistics, Australian National University, Canberra, ACT,
Australia, email: Ross.Maller@anu.edu.au}
\and
Alexander Szimayer\thanks{Department of Business Administration, Universit\"at Hamburg, 20146 Hamburg, Germany, email: Alexander.Szimayer@wiso.uni-hamburg.de}
}

\usepackage{amsmath,latexsym,amssymb,amsthm}
\usepackage{graphicx}	
\usepackage{color}													
\usepackage{multirow}
\usepackage{rotating}

\newcommand{\eqd}{\stackrel{{\cal D}}{=}}
\newcommand{\halmos}{\quad\hfill\mbox{$\Box$}}
\newcommand{\rmd}{{\rm d}}
\newcommand{\rmi}{{\rm i}}
\newcommand{\dirac}{\boldsymbol{\delta}}
\newcommand{\RR}{\mathbb{R}}
\newcommand{\NN}{\mathbb{N}}

\newcommand{\CC}{\mathbb{C}}
\newcommand{\mySS}{\mathbb{S}}
\newcommand{\AAA}{{\cal A}}
\newcommand{\BBB}{{\cal B}}
\newcommand{\CCC}{{\cal C}}
\newcommand{\DDD}{{\cal D}}
\newcommand{\FFF}{{\cal F}}
\newcommand{\GGG}{{\cal G}}

\newcommand{\KKK}{{\cal K}}

\newcommand{\LLL}{{\cal L}}
\newcommand{\MMM}{{\cal M}}
\newcommand{\SSS}{{\cal S}}
\newcommand{\TTT}{{\cal T}}
\newcommand{\OOO}{{\cal O}}

\newcommand{\VVV}{{\cal V}}
\newcommand{\GGC}{GGC}
\newcommand{\VGGC}{VGG}
\newcommand{\skal}[2]{\left\langle #1,#2\right\rangle}
\newcommand{\eins}{{\bf 1}}
\newcommand{\eeee}{\mathfrak{e}}
\newcommand{\rank}{\mbox{rank}}

\newcommand{\diag}{\mbox{diag}}

\newcommand{\Sd}{\mySS^d}
\newcommand{\Sdplus}{\mySS^d_{+}}

\newcommand{\BBarrow}[1]{
\mathrel{\rotatebox[origin=c]{#1}{$\boldsymbol{-\!\!\!-\!\!\!\blacktriangleright}
$}}}

\newtheorem{theorem}{Theorem}
\newtheorem{lemma}{Lemma}
\newtheorem{remark0}{\sc Remark}
\newenvironment{remark}{\begin{remark0}\em}{\end{remark0}\par}

\newtheorem{proposition}{Proposition}
\numberwithin{equation}{section}
\numberwithin{theorem}{section}
\numberwithin{proposition}{section}
\numberwithin{lemma}{section}
\numberwithin{corollary}{section}
\numberwithin{remark0}{section}
\numberwithin{definition}{section}

\begin{document}
\maketitle
\begin{abstract}
We unify and extend a number of approaches related to constructing
multivariate  Madan-Seneta Variance-Gamma model for option pricing.
Complementing Grigelionis' (2007) class, an overarching model is derived by subordinating
multivariate Brownian motion to
a subordinator from Thorin's~(1977) class of generalised Gamma convolutions.
Multivariate classes developed by P\'erez-Abreu and Stelzer~(2014), Semeraro~(2008) and Guillaume~(2013) are submodels.
The classes are shown to be invariant under Esscher transforms, and quite
explicit expressions for canonical measures are obtained, which permit applications such as option pricing using PIDEs or
tree based methodologies. We illustrate with best-of and worst-of
European and American options on two assets.
\end{abstract}
\noindent
\begin{tabbing}
{\em 2000 MSC Subject Classifications:} \ Primary: 60G51,
60F15, 60F17, 60F05\\
\ Secondary: 60J65, 60J75
\end{tabbing}
\vspace{1cm}
\noindent {\em Keywords:} L\'evy Process, Variance-Gamma, Multivariate Subordination, Generalised Gamma Convolutions, Thorin Measure, Esscher Transformation, Esscher Invariance, Superposition, Option Pricing.

\section{Introduction}\label{s1}
Madan and Seneta~\cite{MaSe90} introduced the univariate ``Variance Gamma'' ($VG$) process as a model for a financial asset price process with a special view to more accurate option pricing on the asset, beyond the standard geometric Brownian motion ($GBM$) model. The $VG$ model has proved to be outstandingly successful in this application, and is in common use by many financial institutions, as an alternative to the $GBM$ model, despite failing to cater for stochastic volatility or a leverage effect, or allowing for temporal dependence of
absolute values and/or squares of log returns. Nevertheless, Madan and Seneta extended the $VG$ model~\cite{MaSe90} to a multi-asset version, again with a view to important applications in finance (``rainbow options"),
by subordinating a
multivariate Brownian motion with a single univariate Gamma process (also see~\cite{FiS,FSI,FSII,Se10}). This construction
leaves the marginal processes as univariate Variance Gamma processes. But, components are dependent by
virtue of the common time change.

Semeraro~\cite{Se08} generalises the multi-asset version of Madan and Seneta~\cite{MaSe90} to allow for multivariate subordination.
This permits the dependence structure between asset prices to be modeled in a more flexible way.
The economic intuition behind multivariate subordination is that each asset may have an idiosyncratic risk with its own activity time and a common risk factor, with a joint activity time for all assets. In specific cases it is possible to maintain $VG$ processes for each single asset,
see~\cite{Se08} and related applications in Luciano and Semeraro \cite{LS10,LS10a,LS10b}, though this may be sacrificed for more flexible dependence modeling, as in Guillaume~\cite{Gu12}. Ballotta and Bonfiglioni~\cite{BB14} give an up-to-date account of modeling based on
L\'evy processes in finance.

To summarize, a wide range of multi-asset models based on univariate or multivariate Gamma subordination of a Brownian motion has been proposed. However, there are still gaps in the literature concerning the characterization in general of classes of processes when the class is required to be stable under convolution. Further, theoretical results such as formulae for characteristic functions,  L\'evy measures and, when possible, transition densities, are needed for a  comprehensive description of key properties.

As a prominent link between the real-world and the risk-neutral measure so as to obtain an equivalent martingale measure, we advocate the Esscher transform.
This approach has been rigorously investigated in the seminal papers of Kallsen and
Shiryaev~\cite{KS02} and Eberlein, Papapantoleon and Shiryaev~\cite{EPS09}. Esche and Schweizer~\cite{ES05} show that the minimal entropy martingale measure for a multivariate L\'evy process preserves
the L\'evy property because it corresponds to an Esscher transform. R\"uschendorf and Wolf~\cite{RW15}
provide explicit necessary conditions for the existence of Esscher parameters for multivariate
L\'evy processes and further show that the multivariate Esscher parameter is unique if it exists.
Tankov~\cite{Ta11} provides an introduction to the pricing theory in the context of exponential L\'evy processes.

The aim of the present paper is to contribute to filling the above gaps by presenting a general class of $\RR^d$-valued stochastic processes, constructed by subordinating multivariate Brownian motion with a subordinator drawn from a suitable class of multivariate subordinators. Our intention is to lay out a systematic formulation suitable for future development. For the new processes, we provide the formulae mentioned in the previous paragraph and link the real world and pricing measures by calculating Esscher transforms. To illustrate the practical possibilities, we show how the explicit formulae can be used to price American and European multi-asset options.

\begin{center}
\begin{figure}[hpb!]
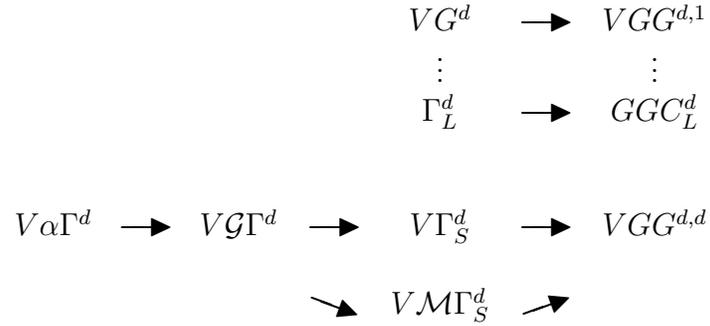

\[\begin{array}{ccccccc}
&&&&VG^d&\BBarrow{000}&\VGGC^{d,1}\\
&&&&\vdots&&\vdots\\
&&&&\Gamma_L^d&\BBarrow{000}&\GGC^{d}_L\\
&&&&&&\\\\
V\alpha\Gamma^d&
\BBarrow{000}&
V\GGG\Gamma^d&
\BBarrow{000}&V\Gamma_S^d&\BBarrow{000}&\VGGC^{d,d}\\\\
&&&\BBarrow{340}&V\MMM\Gamma_S^d&\BBarrow{20}&
\end{array}
\]
\caption{\it Relations between classes of multivariate L\'evy-processes.~Madan-Seneta's~VG~\cite{MaSe90} occurs as a subclass of Grigelionis'\cite{Gr07} $\VGGC^{d,1}$-class;
$\Gamma_L^d\subseteq\GGC_L^d$ Gamma processes and their associated generalised convolution processes based on P\'erez-Abreu and Stelzer~\cite{PS14};
Semeraro's~\cite{Se08}~$V\alpha\Gamma^d$-class, Guillaume's $V\GGG\Gamma^d$-class~\cite{Gu13}, $V\Gamma^d_S=$ Variance Gamma process based on P\'erez-Abreu and Stelzer~\cite{PS14}'s multivariate Gamma subordinators, $\VGGC^{d,d}$-class based on based on P\'erez-Abreu and Stelzer~\cite{PS14}'s $GGC^d$-subordinators; $V\MMM\Gamma^d=$ Variance Matrix Gamma (finitely supported Thorin measures). $-\!\!\!-\!\!\!\blacktriangleright$ points in the direction of generalisation;
$\cdots$~indicates inclusion in special cases.}
\label{diagram}
\end{figure}
\end{center}

The most general class of subordinators we consider is Thorin's~\cite{Th77a,Th77b} class of {\em generalised Gamma convolutions}. We call it the $\GGC$ class of subordinators, and the process formed by subordinating Brownian motion in $\RR^d$ with such a process we call a {\em Variance Generalised Gamma convolution} ($\VGGC$) process.
Grigelionis~\cite{Gr07} constructed such a VGG-class, which we call $\VGGC^{d,1}$ in the present paper. The $\VGGC^{d,1}$ class contains
Madan-Seneta's $VG$ as well as the multivariate generalised hyperbolic
distributions~\cite{Ba77} as special cases. Complementing Grigelionis' $\VGGC^{d,1}$ class, we introduce the $\VGGC^{d,d}$ class of L\'evy processes. Our $\VGGC^{d,d}$ class includes a variety of previously derived models such as Semeraro's $\alpha$-processes~\cite{Se08} and Guillaume's process~\cite{Gu13}. The general $\VGGC=\VGGC^{d,1}\cup \VGGC^{d,d}$ class extends the $VG$ classes in a number of ways. In particular, the $\VGGC$ classes allow for infinite variation and heavy tails.
Figure~\ref{diagram} depicts the connections between the various subordinated classes.
The Thorin~\cite{Th77a,Th77b} generalized Gamma convolutions provide a very
natural class of distributions on which to base our multivariate $VG$
generalizations. As we will show, they facilitate construction of a very
general class of subordinators and corresponding multivariate L\'evy
processes obtained as subordinated $d$--dimensional Brownian motions.
Our new class complements~\cite{Gr07}, and contains a number of currently known versions of multivariate $VG$
processes, and extends them significantly in a variety of important
ways. Although rather technical in appearance, our approach is very much
directed toward practical usage of the methodology. Explicit expressions
for characteristic functions or Laplace transforms, and L\'evy measures
or densities, are derived and exhibited for all our processes. This
permits easy programming of option pricing routines as we demonstrate by
an example, focusing in particular on the pricing of American style
options on a bivariate underlying; a thorny problem not often tackled in
this context.

The $\VGGC$-class and some of its subclasses are shown to be invariant under Esscher transformation, so the risk-neutral distribution constructed as the Esscher transformation of a particular member is also in the  $\VGGC$-class with the practical implication that it is possible to calculate prices under the risk-neutral
measure and estimate model parameters by calibrating to market data.
Using these concepts, we set up a market model and show how an option based on multiple assets may be priced. For illustration we restrict ourselves in this respect to a further subclass of the $\VGGC$-class which we term the $V\MMM\Gamma^d$-processes. Models in this class have the virtue of allowing a quite general dependency structure between the coordinate processes. As an example, we price best--of and worst--of European and American put options, using a tree-based algorithm.

The paper is organised as follows. Section~\ref{secVG} contains the theory. In Subsection~\ref{subsecVGGC}, we revise the Madan-Seneta $VG$ model and set out two major extensions: the Variance-Univariate~$GGC$ and the Variance-Multivariate~$GGC$ classes. This necessitates recalling, first, some basic facts about Gamma subordinators, and then outlining Thorin's
{\em GGC}-class. Subsection~\ref{subsecmoments} gives some results on the moments and sample paths of the new processes, including a computation of Blumenthal-Getoor indices.
In Subsection~\ref{subsecaltstelzerabreu} we investigate relations between our $VGG$-classes and various
multivariate classes of infinitely divisible distributions, as introduced in
P\'erez-Abreu and Stelzer~\cite{PS14}, which we call $\Gamma_L^d\subseteq\GGC^d_L$-classes. Subsection~\ref{subsecEsscher} calculates Esscher transforms, stating the fact that both~$VGG$-classes remain invariant. Subsection~\ref{subsecVGamma} introduces the Variance-$\MMM\Gamma^d$-subclass on which we base the option pricing model in Section~\ref{secapplic}. Finally, Subsection~\ref{subsecaltsubordclass} collects further properties of our subordinator class, including comparisons with those occurring in the literature.

Section~\ref{secapplic} contains applications. Here the market model is introduced, risk-neutral valuation is discussed, and in Subsection~\ref{subsecoption} we price
some cross-dependence sensitive options of both European and American types.
Some illustrations of the kinds of dependencies the models allow are also given there. Proofs of Subsections~\ref{subsecVGGC}--\ref{subsecEsscher} and necessary methodological tools are relegated to Section~\ref{secproof}, where polar decomposition of measures and subordination are briefly covered.
\section{Theory}\label{secVG}
\subsection{Variance Generalised Gamma Convolutions ($\VGGC$)}\label{subsecVGGC}
\noindent{\bf Preliminaries.}
$\RR^d$ is the $d$-dimensional Euclidean space; elements of $\RR^d$ are column vectors $x=(x_1,\dots,x_d)'$. Let $\skal xy$ denote the Euclidean product
and $\|\cdot\|_E^2:=\skal xx$ be the Euclidean norm, and set $\|x\|^2_\Sigma:=\skal{x}{\Sigma x}$ for $x,y\in\RR^d$ and $\Sigma\in\RR^{d\times d}$. For $A\subseteq\RR^d$ let $A_*=A\backslash\{0\}$.
$\eins_A$ denotes the indicator function. The Dirac measure with total mass at $x\in\RR^d$ is $\dirac_x$. $I:\RR\to\RR$ denotes the identity function.

$X=(X_1,\dots,X_d)'=(X(t))_{t\ge 0}$ is a $d$-dimensional
L\'evy
process if $X$ has independent and stationary increments, $X(0)=0$ and the sample
paths
$t\mapsto X(t)\in\RR^d$ are c\`adl\`ag functions.

The law of a L\'evy process $X$ is determined by its characteristic function via $Ee^{\rmi\skal\theta{X(t)}}=\exp\{t\psi_X(\theta)\}$ with L\'evy exponent, for $t\ge 0$, $\theta\in\RR^d$,
\begin{equation}\label{0.1}
\psi_X(\theta)\,=\,
\rmi \skal {\gamma_X}\theta\!-\!\frac 12\;\|\theta\|^2_{\Sigma_X}
+\int_{\RR_*^d}\left(e^{\rmi\skal\theta x}\!-\!1\!-\!\rmi\skal\theta x \eins_{\|x\|\le
1}\right)\,\Pi_X(\rmd x)\,.
\end{equation}
Here $\gamma_X\in\RR^d$, $\Sigma_X\in\RR^{d\times d}$ is a symmetric nonnegative definite matrix,
$\Pi_X$ is a nonnegative Borel measure on $\RR^d_*$ satisfying
\begin{equation}\label{Piintegrab}
\int_{\RR^d_*}\,\|x\|^2\wedge 1\;\Pi_X(\rmd x)\;<\;\infty\,,
\end{equation}
and $\|\cdot\|$ is a given norm on $\RR^d$.
We write $X\sim L^d(\gamma_X,\Sigma_X,\Pi_X)$ whenever $X$ is a $d$-dimensional L\'evy process with canonical triplet
$(\gamma_X,\Sigma_X,\Pi_X)$; $BM^d(\gamma,\Sigma):=L^d(\gamma,\Sigma,0)$ refers to Brownian motion with drift $\gamma$ and covariance matrix $\Sigma$.

Paths of $X$ are of (locally) {\em finite variation ($FV^d$)} whenever $\Sigma_X=0$ and
\begin{equation}\label{PiFVintegrab}
\int_{0<\|x\|\le 1}\|x\|\,\Pi_X(\rmd x)\quad<\quad\infty\,.
\end{equation}
In this case, we write $X\sim FV^d(D_X,\Pi_X)$ with $D_X$ denoting the drift of $X$: $D_X:=\gamma_X-\int_{0<\|x\|\le 1} x\;\Pi_X(\rmd x)\;\in\;\RR^d$.
A $d$-dimen\-sional L\'evy process $T$ with nondecreasing components is called a $d$-dimensional {\em subordinator}, possibly with drift $D_T$, written $T\sim S^d(D_T,\Pi_T)$. A general L\'evy process $X\sim L^d(\gamma_X,\Sigma_X,\Pi_X)$ is a subordinator with drift $D_X$ if and only if  $X\sim FV^d(D_X,\Pi_X)$
with $D_X\in[0,\infty)^d$ and $\Pi_X$ being concentrated on $[0,\infty)_*^d$.

We write $X\eqd Y$ and $X\sim Q$ whenever $\LLL(X)=\LLL(Y)$ and $\LLL(X)=Q$, respectively, where $\LLL(X)$ denotes the law of a random variable or stochastic process $X$. There is a correspondence between infinitely divisible distributions and L\'evy processes $X$: for all $t\ge 0$ the law of $X(t)$, $P(X(t)\in\rmd x)$, is infinitely divisible. Vice versa, any infinitely divisible Borel probability measure $Q$ on $\RR^d$ determines uniquely the distribution of a L\'evy process via $X(1)\sim Q$. This connection is used throughout the
paper. For instance, we write $T\sim Q_S$ to indicate that $T$ is a subordinator with $T(1)\sim Q$. See~\cite{A,b,CS09,CT,KSW05,s} for basic properties of L\'evy processes and their applications in finance.\\[1mm]
{\bf Subordination.} In~\cite{BPS01} various kinds of subordination are introduced~(see Subsection~\ref{subsecsubord} for details).
In the present paper, we will make use of two extreme cases: univariate and (strictly) multivariate subordination. Let $X=(X_1,\dots,X_d)'$ be a $d$-dimensional L\'evy process. $X$ serves as the subordinate.

Given a univariate subordinator $T$, independent of $X$, define a $d$-dimen\-sional L\'evy process, denoted $X\circ T$, by setting
\begin{equation}\label{defunisub}(X\circ T)(t):=(X_1(T(t)),\dots,X_d(T(t)))',\qquad t\ge 0\,.\end{equation}
In the sequel, we denote the law of $X\circ T$ by $\LLL(X)\circ \LLL(T)$.
We refer to this type as {\em univariate subordination} (cf.~Section~6 in \cite{s}).

Suppose $X$ has {\em independent components} $X_1,\dots,X_d$.  Let $T=(T_1,\dots,T_d)'$ be a
$d$-dimensional subordinator, independent of $X$, and define a $d$-dimensional L\'evy process by setting
\begin{equation}\label{defmultisubinproof}X\,\circ_{d}\, \,T:=(X_1\circ T_1,\dots,X_d\circ T_d)'\,.\end{equation}
The law of $X\circ_{d} T$ is denoted by $\LLL(X)\,\circ_{d}\, \LLL(T)$.
\begin{remark}\label{countermultivarexample}
When dealing with strictly multivariate subordination, we have to restrict the class of admissible subordinates $X$ to L\'evy processes with {\em independent components}. This is necessary if we are to stay in the class of L\'evy processes. For instance, let
$B\sim BM_1(0,1)$ be a univariate standard BM. Then $X=(B,B)'$ is a L\'evy process, but $t\mapsto(B(t),B(2t))'$ is not.
\halmos
\end{remark}
\noindent{\bf Gamma subordinator.} Denote by $\Gamma(\alpha,\beta)$ a Gamma distribution with parameters $\alpha,\beta>0$, i.e.,
a Borel probability measure on $\RR$, absolutely continuous with respect to Lebesgue measure, defined by
\begin{equation}\label{defgamma}\Gamma(\alpha,\beta)(\rmd x)\,=\,
\eins_{(0,\infty)}(x)\;\frac{\beta^{\;\alpha}}{\Gamma(\alpha)}\; x^{\alpha-1} e^{-\beta x}\,\rmd x\,,\qquad x\in\RR\,.\end{equation}
We write $G\sim \Gamma_S(\alpha,\beta)$ for a {\em Gamma process} $G=(G(t))_{t\ge 0}$ with parameters $\alpha$, $\beta>0$, that is, $G$ is a univariate subordinator having marginal distributions $G(t)\sim \Gamma(\alpha t,\beta)$, $t>0$. Further, for $\lambda>-\beta$, $t>0$, recall that
\begin{eqnarray}\label{laplacesubgamma}Ee^{-\lambda G(t)}=\left\{(\beta\big/(\beta+\lambda)\right\}^{\alpha t}
= \exp\Big\{-t\,\int_0^\infty \left(1-e^{-\lambda r}\right)\,
\alpha e^{-\beta r}\;\frac{\rmd r}r\Big\}
\end{eqnarray}
(the first formula is well known, the second identity is Frullani's integral, see~p.16 \& p.73~in \cite{b}). In particular, a Gamma process has zero drift, and its L\'evy measure admits a Lebesgue density
$\Pi_G(\rmd r)\,=\,
\eins_{(0,\infty)}(r)\;\alpha e^{-\beta r}\,\rmd r\big/r$.

For $\alpha=\beta$ we have $E[G(1)]=1$. $G$ is then also called a {\em standard Gamma process}, briefly $G\sim \Gamma_S(\alpha):=\Gamma_S(\alpha,\alpha)$.\\[1mm]
\noindent{\bf Madan-Seneta $\boldsymbol{VG^d}$ Process.} Madan and Seneta~\cite{MaSe90} (for extensive investigations and reviews $cf$.~\cite{FiS,FSI,FSII,
KT08,KT09,MaCaCh,Se10}) suggest subordinating Brownian motion with a Gamma process. For the parameters of this model we assume $\mu\in\RR^d$, $b>0$ and $\Sigma\in\RR^{d\times d}$, with $\Sigma$ being symmetric and nonnegative definite.

Let $B\sim BM^d(\mu,\Sigma)$ be a $d$-dimensional
Brownian motion and $G\sim \Gamma_S(b)$ be independent of $B$. A L\'evy process $Y$ is a $d$-dimensional {\em Variance Gamma $(VG^d)$}-process with parameters $b,\mu,\Sigma$ whenever $Y\eqd B\circ G$, which we write as
\begin{equation}\label{defMSVG}Y\sim VG^d(b,\mu,\Sigma):=BM^d(\mu,\Sigma)\circ \Gamma_S(b)\,.\end{equation}
(a)~Note that a $VG^d$-process has zero drift and is of finite variation.\\[1mm]
(b)~The Laplace transformation of $Y$ takes on an explicit form, straight-forwardly derived from conditioning: for $t\!\ge\!0$, $\lambda\!\in\!\RR^d$ with $\frac 12\|\lambda\|^2_\Sigma-\skal\mu \lambda<b$,
\begin{eqnarray}
E\exp\{-\skal \lambda {Y(t)}\}
&=& \Big\{b\Big/(b+\skal\mu\lambda-\frac 12\|\lambda\|^2_{\Sigma})\Big\}^{b t}\,.\label{VGLaplace}
\end{eqnarray}
\noindent (c)~If $\Sigma$ is invertible, for $t>0$ the distribution of $Y(t)$ is absolutely continuous with respect to the Lebesgue measure with the following density:
\begin{eqnarray}\nonumber
f_{Y(t)}(y)&=&
\frac{2^{(2-d)/2}\;b^{b t}\;\exp\{\skal{\Sigma^{-1}\mu}{y}\}}{\pi^{d/2}\;(\det\Sigma)^{1/2}\;\Gamma(b t)}\left\{\frac{\|y\|^2_{\Sigma^{-1}}}{2b\!+\!\|\mu\|^2_{\Sigma^{-1}}}\right\}^{(2b t-d)/4}\!\!\!\!\!\!\!\!\!\times\\
&&{}\times K_{|2b t-d|/2}\left(\sqrt{(2b\!+\!\|\mu\|^2_{\Sigma^{-1}})\;\|y\|^2_{\Sigma^{-1}}}\right),\quad y\!\in\!\RR^d\,.\label{VGlaw}
\end{eqnarray}
Further, still with $\det \Sigma\neq 0$, the canonical L\'evy measure of $Y$ is absolutely continuous with respect to
Lebesgue measure and satisfies:
\begin{eqnarray}\nonumber
\frac{\rmd \Pi_Y}{\rmd y}(y)&=&\frac{b \;2^{(2-d)/2}\;\exp\{\skal{\Sigma^{-1}\mu} y\}}{\pi^{d/2}\;(\det\Sigma)^{ 1/2}}\left\{\frac{2b\!+\!\|\mu\|^2_{\Sigma^{-1}}}{\|y\|^2_{\Sigma^{-1}}}\right\}^{d/4}\times\\
&&{}\times\;K_{d/2}\left(\sqrt{(2b\!+\!\|\mu\|^2_{\Sigma^{-1}})\;\|y\|^2_{\Sigma^{-1}}}\right), \quad
y\in\RR^d.\label{VGlev}
\end{eqnarray}
Here $K_\nu$ is the {\em modified Bessel function of the second kind} (see~\cite{GrRy96}, their Equations~(3.471)--9 and~(8.469)--3;
also~\cite{CT}, their Appendix). It is convenient to revise the following
facts about the modified Bessel function $K_\nu$ of the second kind: for $\delta,\gamma>0$, $\nu\in\RR$
\begin{equation}
2\left(\delta/\gamma\right)^{\nu/2}\,K_{|\nu|}\big(2\sqrt{\delta\gamma}\big)\,=\,\int_0^\infty r^{\nu-1}\,
\exp\big\{-(\delta/r)-\gamma r\big\}\,\rmd r\,.\label{Bessel}
\end{equation}
For large values of $x$~(see~\cite{GrRy96}, their  Equation~(8.451)--6),
\begin{equation}\label{Besselass}
K_{\nu}(x)\sim K_{1/2}(x)=\sqrt{\pi}\;e^{-x}/\sqrt{2x}\,,\quad x\to\infty\,.
\end{equation}
We use a variant of $K_\nu$, defined by
\begin{equation}\label{varmodBess}
\widehat K_\nu(r):=r^\nu K_\nu(r)\,,\qquad r,\nu>0\,.
\end{equation}
\begin{remark}\label{remGGCexplan} We get from~\eqref{laplacesubgamma} that the shifted Gamma process $\widetilde G(t)= a t + G(t)$, $t\ge 0$, with $G\sim \Gamma_S(\alpha,\beta)$ has Laplace transform for $\lambda\ge 0$
\[
-\log E e^{-\lambda\widetilde G(t)}=\lambda a t+\alpha t \log \frac{\beta+\lambda}\beta
=\lambda a t+t \int_{(0,\infty)}\log (1+(\lambda/y))\,\alpha \dirac_\beta(\rmd y)\,.
\]
For independent processes $G_1,\dots,G_n$ with $G_k=\Gamma_S(\alpha_k,\beta_k)$, $1\le k\le n$, introduce the associated shifted Gamma processes by $\widetilde G_k(t)= a_k t + G_k(t)$, $t\ge 0$, $1\!\le\! k\!\le\! n$. It follows from the independence that for $\lambda\ge 0$
\[
-\log E e^{-\lambda \sum_{i=1}^n\widetilde G_i(t)}=-\sum_{k=1}^n\log E e^{-\lambda\widetilde G_i(t)}=t\lambda a+t\int_{(0,\infty)}\!\!\log( 1+(\lambda/y))\,\TTT_n(\rmd y),
\]
where $a=\sum_{k=1}^n a_k$ and $\TTT_n=\sum_{k=1}^n\alpha_k\dirac_{\beta_k}$ is the discrete measure associated with the increasing function $U_n(y)=\sum_{k=1}^n \alpha_k \eins_{[\beta_k,\infty)}(y)$, $y>0$.
Using the representation in the last display and taking suitable limits in distribution, we arrive at the class of generalised Gamma convolutions. In the multivariate setting
in Subsections~\ref{subsecVGamma}--\ref{subsecaltsubordclass}, we shall employ finitely supported Thorin measures, such as $\TTT_n$ above, in the context of the $\MMM\Gamma^d$--class.\halmos
\end{remark}
\noindent{\bf Generalised Gamma Convolution Subordinator~($\boldsymbol{GGC}$).} The class of Gamma distributions is not closed under convolutions. To extend the Madan-Seneta $VG^d$-class it is convenient to use the subordinators corresponding to Thorin's class~\cite{Th77a,Th77b} of generalised Gamma convolutions ($\GGC$). This is the smallest class of distributions that contains all Gamma distributions, but is closed under convolution and weak convergence (see~\cite{Bo92,Bo09,Gr07,JLY08,SSV,SH04}; for multivariate extensions see~\cite{BMS06,Bo09,PS14}). The class of GGC-distributions is a subclass
of the self-decomposable distributions and, thus, the distributions are infinitely divisible.

A $d$-dimensional {\em Thorin measure} $\TTT$ is a Borel measure on $[0,\infty)^d_*$ with
\begin{equation}\label{thorinmeasure}
\int_{[0,\infty)^d_*}\;\left(1\!+\!\log^-\|x\|\right)\wedge\left(1\big/\|x\|\right)\;\TTT(\rmd x)\quad <\quad \infty\,.
\end{equation}
(Throughout $x=x^+-x^-$ denotes the decomposition of an extended real number $x\in\overline\RR$ into positive and negative parts.)

A subordinator $T$ is a {\em $\GGC^d$-subordinator with parameters}\ $a$ and $\TTT$, in brief $T\sim \GGC_S^d(a,\TTT)$,
when $\TTT$ is a $d$-dimensional Thorin measure, $a\in[0,\infty)^d$ and, for all $t\ge0,\lambda\in[0,\infty)^d$,
\begin{equation}\label{LaplaceGGCd}
-\log E\exp\{-\skal{\lambda}{T(t)}\}\,=\,t\skal a\lambda+t\int_{[0,\infty)^d_*} \log\left\{\frac{\|x\|^2+\skal{\lambda}{x}}{\|x\|^2}\right\}\;\TTT(\rmd x)\,.
\end{equation}
The distribution of a Thorin subordinator is determined by $a$ and $\TTT$. Any Thorin measure $\TTT$
admits a polar representation $\TTT=\alpha\otimes_p \KKK$ relative to $\mySS^d_{+}:=\{x\in[0,\infty)^d:\|x\|=1\}$ (see Lemma~\ref{lempolarrec} below).
This allows us to specify the corresponding L\'evy measure. For $T\sim \GGC^d_S(a,\TTT)$ with $\TTT=\alpha\otimes_p \KKK$ we have
$T\sim S^d(a,\Pi_T)$ with
\begin{eqnarray}\label{GGClevydensity}
\rmd\Pi_T&=&\int_{\mySS_+^d}\int_0^\infty\dirac_{rs}\,k(s,r)\,\frac{\rmd r}r\;\alpha(\rmd s)\,,\\
k(s,r)&=&\int_{(0,\infty)} e^{-r \tau}\,\KKK(s,\rmd \tau)\,,\;\quad r>0,s\in\Sdplus\,,\label{GGCkfunction}
\end{eqnarray}
see~\cite{BMS06,PS14}, their Theorem~F and Proposition~4.3, respectively.\\[1mm]
\noindent{\bf Variance--Univariate GGC $(\boldsymbol{\VGGC^{d,1}})$.} As a first extension of the $VG^d$-model, we review Grigelionis'~\cite{Gr07} class. Grigelionis used univariate subordination $\circ$ and subordinated Brownian motion with a univariate $\GGC$-subordinator. For the parameters of his model we take $\mu\in\RR^d$, $a>0$ and $\Sigma\in\RR^{d\times d}$, with $\Sigma$ being symmetric and nonnegative definite. Further, let $\TTT$ be a univariate Thorin measure. Let $B\sim BM^d(\mu,\Sigma)$ be a $d$-dimensional
Brownian motion, $T\sim \GGC^1_S(a,\TTT)$, independent of $B$.

Given such $B$ and $T$, we call a L\'evy process of the form $Y\eqd B\circ~T$ a $d$-dimen\-sional {\em Variance Univariate Generalised Gamma Convolution $($$\VGGC^{d,1}$$)$}-process with parameters $a,\mu,\Sigma,\TTT$. We write this as
\begin{equation}\label{defGVGd1}Y\sim \VGGC^{d,1}(a,\mu,\Sigma,\TTT):=BM^d(\mu,\Sigma)\circ \GGC^1_S(a,\TTT)\,.\end{equation}
The next theorem gives the characteristic function and L\'evy density.
Part~(a) (our Equation~\eqref{GVGlevdensI}) is proved in Subsection~\ref{subsecproofVGGC}.
Part~(b) occurs in \cite{Gr07} (see his Proposition~3). (In \eqref{charGVG}, $\log:\CC\backslash(-\infty,0]\to\CC$ denotes the principal branch of the logarithm.)
\begin{theorem}\label{theoGVG} Let $Y\sim \VGGC^{d,1}(a,\mu,\Sigma,\TTT)$.\\[1mm]
(a)~For all $\theta\in\RR^d$, $t\ge 0$, we have $E\exp\{\rmi \skal\theta {Y(t)}=\exp\{t\psi_Y(\theta)\}$ with
\begin{equation}\label{charGVG}
\psi_Y(\theta)=\rmi a \skal\mu\theta-(a/2)\|\theta\|^2_\Sigma-
\int_{(0,\infty)}\!\!\log\frac{\tau\!-\!\rmi\skal\mu\theta\!+\!\frac 12\|\theta\|^2_\Sigma}\tau\,\TTT(\rmd\tau)\,.
\end{equation}
(b)~Assume~$\det\Sigma\neq 0$ and $\TTT\neq 0$. Then $\Pi_Y$ is absolutely continuous with respect to
$d$-dimensional Lebesgue measure on $\RR^d_*$, where, for $y\in\RR^d_*$,
\begin{eqnarray}
\label{GVGlevdensI}
\lefteqn{\frac{\rmd \Pi_Y}{\rmd y}(y)
\,=\,2^{(2-d)/2}\;\pi^
{-d/2}\;(\det\Sigma)^{-1/2}\;\|y\|^{-d/2}_{\Sigma^{-1}}\;\exp\{\skal{\Sigma^{-1}\mu} y\}\;\times}&&\\
&&{}\times\int_{(0,\infty)}
(2\tau\!+\!\|\mu\|^2_{\Sigma^{-1}})^{d/4}\;K_{d/2}\Big(\sqrt{(2\tau\!+\!\|\mu\|^2_{\Sigma^{-1}})\;\|y\|^2_{\Sigma^{-1}}}\Big)\,
\TTT(\rmd \tau)\,.\nonumber
\end{eqnarray}
Besides this, we have $\Pi_Y=\int_{\mySS^d_E} \left(\int_0^\infty\dirac_{rs}\; g_Y(s,r)\,\rmd r/r\right)\, \alpha_{d,E}(\rmd s)$
in Euclidean polar coordinates with Lebesgue surface measure $\alpha_{d,E}$ on $\mySS^d_E$ (for $d=1$ we interpret $\alpha_{d,E}$ as the counting measure), for $r>0$, $s\in \mySS_E$,
\begin{eqnarray}
\label{GVGlevdensII}
\lefteqn{\hspace*{-5em}
g_Y(s,r)
\,=\,2^{(2-d)/2}\;\pi^
{-d/2}\;(\det\Sigma)^{-1/2}\;\|s\|^{-d}_{\Sigma^{-1}}\;\exp\{r\skal{\Sigma^{-1}\mu} s\}\;\times}&&\\
&&{}\times\int_{(0,\infty)}\widehat K_{d/2}\Big((2\tau\!+\!\|\mu\|^2_{\Sigma^{-1}})^{1/2} \; r\|s\|_{\Sigma^{-1}}\Big)\,
\TTT(\rmd \tau)\,.\nonumber
\end{eqnarray}
\end{theorem}
\begin{remark}\label{remVGGCdvs1}
In both
classes, $VG^d$ and $\VGGC^{d,1}$, we subordinate a Brownian motion with  a single univariate subordinator.
Thus the components of these processes must
jump simultaneously. To allow the components to jump independently of each other we must use multivariate subordination of Brownian motion. This motivates our next step, the introduction of our $\VGGC^{d,d}$-class.~\halmos\end{remark}
\noindent{\bf Variance--Multivariate GGC $(\boldsymbol{\VGGC^{d,d}})$.}~Next we give another modification of the $VG^d$-model which is constructed by multivariate subordination~$\circ_{d}$.
For the parameters of this model we assume a $d$-dimensional Thorin measure $\TTT$, $\mu\in\RR^d$, $a\in[0,\infty)^d$ and $\Sigma\in\RR^{d\times d}$, with $\Sigma=\diag(\Sigma_{11},\dots,\Sigma_{dd})$ having nonnegative entries. (We impose on Brownian motion the requirement to have independent components, so as to stay in the class of L\'evy processes, see Remark~\ref{countermultivarexample}.)

Let $B\sim BM^d(\mu,\Sigma)$ be a Brownian motion.
Let $T\sim \GGC_S^d(a,\TTT)$ be independent of $B$. Given such $B$ and $T$, we call a L\'evy process of the form $Y\eqd B\circ_{d}~T$ a $d$-dimensional {\em Variance Multivariate Generalised Gamma Convolution $($$\VGGC^{d,d}$$)$}-process with parameters $a,\mu,\Sigma,\TTT$. We write this as
\begin{equation}\label{defGVG}Y\sim \VGGC^{d,d}(a,\mu,\Sigma,\TTT):=BM^d(\mu,\Sigma)\circ_{d} \GGC^d_S(a,\TTT)\,.\end{equation}
To state formulae for the characteristics of this process, define the outer $\diamond$-products of $y=(y_1,\dots,y_d)',z=(z_1,\dots,z_d)'\in\RR^d$ and $\Sigma\in\RR^{d\times d}$
as
\begin{eqnarray}
y\diamond z&:=&(y_1z_1,y_2z_2,\dots,y_dz_d)'\in\RR^d\,,\label{defotimes1}\\
\Sigma\diamond z&:=&\diag(z_1,\dots,z_d)\Sigma\in\RR^{d\times d}\,.\nonumber
\end{eqnarray}
We can decompose $[0,\infty)^d_*=\bigcup_{\emptyset\neq J\subseteq \{1,\dots,d\}} C_J$ into semi-cones, where
\begin{equation}\label{coneJ} C_J:=\Big\{\sum_{j\in J} x_j \eeee_j:x_j>0\mbox{ for all $j\in J$}\Big\}\,, \qquad \emptyset \neq J\subseteq\{1,\dots,d\}\,,\end{equation}
and $\eeee_i$ are the unit coordinate vectors. Analogously, we can decompose $\RR^d_*=\bigcup_{\emptyset\neq J\subseteq \{1,\dots,d\}} V_J$ into $V_J:=\{\sum_{j\in J} x_j \eeee_j:x_j\neq 0\mbox{ for all $j\in J$}\big\}$ for $\emptyset \neq J\subseteq\{1,\dots,d\}$. Let $\#J$ be the cardinality of~$J$.

We need a family of reference measures. With $\ell$ denoting the univariate Lebesgue measure define $\ell_J:=\bigotimes_{k=1}^d \ell_{J,k}$ as the product measure with the following factors:
\begin{equation}\label{deflambdaJk}
\ell_{J,k}\,:=\,\eins_{J}(k)\ell+\eins_{\{1,\dots,d\}\backslash J}(k)\dirac_{\bf0}\,,\qquad 1\le k\le d\,.
\end{equation}
Observe $\ell_{J}(\RR^d_*-V_J)=0$.
Finally, to provide an analog of~\eqref{GVGlevdensII} in Euclidean polar coordinates, let $\alpha_{J,E}$
denote the Lebesgue surface measure on $\mySS^d_E \cap V_J$ for $J\neq\emptyset$. (In the discrete case we interpret
$\alpha_{J,E}$ as counting measure).

The next theorem gives the characteristic function of $Y$ and an expression for its L\'evy measure. It is proved in Subsection~\ref{subsecproofVGGC}.
\begin{theorem}\label{theoGVGmulti}Let $Y\sim \VGGC^{d,d}(a,\mu,\Sigma,\TTT)$.\\[1mm]
(a)~For all $\theta\in\RR^d$, $t\ge 0$, we have $E\exp\{\rmi \skal\theta {Y(t)}\}=\exp\{t\psi_Y(\theta)\}$ with
\begin{equation}\label{charGVGmulti}
\psi_Y(\theta)=
\rmi\skal{\mu\diamond a}\theta-\frac 12\|\theta\|^2_{\Sigma\diamond a}
-\int_{[0,\infty)^d_*}\!\!\log\frac{\|x\|^2-\rmi \skal {\mu\diamond x} \theta+\frac 12\|\theta\|^2_{\Sigma\diamond x}}{\|x\|^2}\,\TTT(\rmd x)\,.
\end{equation}
(b)~Always, $\Pi_Y=\sum_{\emptyset\neq J\subseteq \{1,\dots,d\}}\Pi_J$, with $\Pi_J(\RR^d_*-V_J)=0$ for $\emptyset\neq J\subseteq \{1,\dots,d\}$.
If $\TTT(C_J)=0$ then $\Pi_J\equiv0$. Otherwise, if  $\TTT(C_J)>0$ and $\det\Sigma>0$ then $\Pi_J$ is absolutely continuous with respect to $\ell_J$ and, for $y\in V_J$,
\begin{eqnarray}
\label{GVGlevdensmultiI}
\frac{\rmd \Pi_J}{\rmd \ell_J}(y)&=&2^{(2-\#J)/2}\;\pi^
{-\#J/2}\;\exp\{\skal{\Sigma^{-1}\mu}y\}\times\\
&&{}\int_{C_J}\,\frac{\TTT(\rmd x)}{\prod_{j\in J}\Sigma_{jj}^{1/2}x_j^{1/2}}
\;\;\;\left\{\frac{2\|x\|^2\!+\!\skal{\mu\diamond x}{\Sigma^{-1}\mu}}{\sum_{j\in J}y_j^2/(x_j\Sigma_{jj})}\right\}^{\#J/4}\;\times\nonumber\\
&&{}\qquad K_{\#J/2}\left(\bigg\{\Big(2\|x\|^2\!+\!\skal{\mu\diamond x}{\Sigma^{-1}\mu}\Big)\sum_{j\in J}y_j^2/(x_j\Sigma_{jj})\bigg\}^{1/2}\right)\,,\nonumber
\end{eqnarray}
whereas in
$\Pi_J=\int_{\mySS^d_E \cap V_J}(\int_0^\infty\dirac_{rs}\; g_J(s,r)\,\rmd r/r)\alpha_{J,E}(\rmd s)$, for $r>0$, $s\in \mySS^d_E\cap V_J$,
\begin{eqnarray}
\label{GVGlevdensmultiII}
\lefteqn{g_J(s,r)
\,=\,2^{(2-\#J)/2}\;\pi^
{-\#J/2}\;\exp\{r\skal{\Sigma^{-1}\mu} s\}\;\times}&&\\
&&{}\int_{C_J}\,\frac{\TTT(\rmd x)}{\prod_{j\in J}\Sigma_{jj}^{1/2}x_j^{1/2}}
\;\;\;\Big\{\sum_{j\in J}s_j^2/(x_j\Sigma_{jj})\Big\}^{-\#J/2}\;\times\nonumber\\
&&{}\qquad \widehat K_{\#J/2}\Big(r\times \Big\{\big(2\|x\|^2\!+\!\skal{\mu\diamond x}{\Sigma^{-1}\mu}\big)\sum_{j\in J}s_j^2/(x_j\Sigma_{jj})\Big\}^{1/2}\Big)\,.\nonumber
\end{eqnarray}
\end{theorem}
\subsection{Moments and Sample Paths}\label{subsecmoments}
In Subsection~\ref{subsecaltsubordclass}, we see that both $\VGGC$-classes support pure jump processes with infinite variation and infinite moments. In Propositions~\ref{propGGCmoments}--\ref{propVGGCmoments}, we provide conditions on the Thorin measure
that can be used to check local integrability of $\Pi_T$ and $\Pi_Y$ as well as existence of moments (for a proof see Subsection~\ref{subsecproofmoments}). Set $\|\cdot\|_1:=|\cdot|$ and $\|\cdot\|_d:=\|\cdot\|$.
\begin{proposition}\label{propGGCmoments} Let $t>0$, $0\!<\!q\!<\!1$, $p>0$ and $T\sim \GGC^d_S(a,\TTT)$.
Then:\\[2mm]
(a)~\;$\int_{0<\|z\|\le 1} \|z\|^q\;\Pi_T(\rmd z)<\infty\quad\Leftrightarrow\quad\;\int_{\|x\|>1} \TTT(\rmd x)/\|x\|^{q}<\infty$.\\[2mm]
(b)~\quad\quad\quad\quad\;$E[\|T(t)\|^p]<\infty\;\quad\Leftrightarrow\quad\int_{0<\|x\|\le 1} \TTT(\rmd x)/\|x\|^{p}<\infty$.
\end{proposition}
\begin{proposition}\label{propVGGCmoments} Let $k\!\in\!\{1,d\}$ and $Y\!\sim\! \VGGC^{d,k}(a,\mu,\Sigma,\TTT)$.
Then we have:\\[1mm]
(a)~Let $0\!<\!q\!<\!2$. If
\begin{equation}\label{intTTTqhalffinite}
\int_{\|x\|_k>1} \TTT(\rmd x)/\|x\|_k^{q/2}<\infty\,\end{equation}
then
\begin{equation}\label{intPIYqfinite}
\int_{0<\|y\|\le 1}\|y\|^q\;\Pi_Y(\rmd y)<\infty\,.\end{equation}
If, in addition, $\det \Sigma>0$ then~\eqref{intTTTqhalffinite} and~\eqref{intPIYqfinite} are equivalent.
\\[2mm]
(b)~Let $p,t\!>\!0$. If
\begin{equation}\label{intTTTphalffinite}
\left\{\begin{array}{ll}\int_{[0,1]^d_*}\TTT(\rmd x)/\|x\|^{p/2}_k<\infty&\mbox{when $\mu=0$}\\
\int_{[0,1]^d_*}\TTT(\rmd x)/\|x\|^{p}_k<\infty&\mbox{when $\mu\neq 0$}
\end{array}\right.\,\end{equation}
then
\begin{equation}\label{intpmomentYfinite}E[\|Y(t)\|^p]<\infty\,.\end{equation}
If, in addition, either $\prod_{k=1}^d\mu_k\neq 0$, or $\mu=0$ and $\det \Sigma>0$, then~\eqref{intTTTphalffinite} and~\eqref{intpmomentYfinite} are equivalent.
\end{proposition}
\begin{remark}\label{remGIGGGCresavenue} Subordinating Brownian motion with independent subordinators from the generalised inverse Gaussian (GIG)-subordinators
one obtains the class of generalised hyperbolic L\'evy processes~(\cite{Ba77,BS01,Eb01} for detailed accounts).
Halgren~\cite{Ha79} identified the univariate GIG distributions as generalised
Gamma convolutions, with the implication that the associated class of hyperbolic L\'evy processes~\cite{Ba77}
forms a subclass of the~$\VGGC^{d,1}$-class (see~\cite{Bo92} and~\cite{Gr07}, his Example~1).

Owing to Proposition~\ref{propVGGCmoments}, we may restrict our analysis to the subordinator class. For $(\alpha,\beta,\gamma)\in \RR\times(0,\infty)^2\cup
(0,\infty)^2\times\{0\}\cup(-\infty,0)\times\{0\}\times(0,\infty)$ the $GIG(\alpha,\beta,\gamma)$-distribution has the following probability density on $(0,\infty)$
\[GIG(\alpha,\beta,\gamma)(\rmd x)\,=\,C_{\alpha,\beta,\gamma}\,x^{\alpha-1}\,e^{-\beta x-(\gamma/x)}\;\rmd x\,,\quad x>0\,.\]
Here $C_{\alpha,\beta,\gamma}$ is a normalising constant. For $\gamma=0$ the identity $GIG(\alpha,\beta,0)=\Gamma(\alpha,\beta)$ holds,
and exponential moments $E[e^{\lambda T(1)}]$ are finite for $\lambda<\beta$, and this extends to other parameters $\alpha,\beta,\gamma$ as long as $\beta>0$. For $\beta=0$ and $\alpha<0$
observe $GIG(\alpha,0,\gamma)=$inv$\Gamma(-\alpha,\gamma)$ is the inverse Gamma distribution with finite $p$-moments of order $p<|\alpha|$ only.

We determine the Blumenthal-Getoor index~\cite{BG61} as an indicator for activity of the associated $GIG$-subordinator, for potential applications see~\cite{AJ12,TT10}. In~\eqref{laplacesubgamma}, note  $\int_{(0,1]} x^q \Pi_G(\rmd x)=\alpha\int_0^1 x^{q-1} e^{-\beta x}\rmd x$ is finite for all $0\!<\!q\!<\!1$, forcing the
Blumenthal-Getoor index of $GIG(\alpha,\beta,0)=\Gamma(\alpha,\beta)$ to degenerate to zero. For the remaining parameters, where $\gamma>0$, we compute the Laplace transform of $GIG(\alpha,\beta,\gamma)$ for $\lambda>0$ as
\begin{eqnarray*}
\exp\{-\Lambda_{\alpha,\beta,\gamma}(\lambda)\}&:=&\int_0^\infty e^{-\lambda x}GIG(\alpha,\beta,\gamma)(\rmd x)=\frac{C_{\alpha,\beta,\gamma}}{C_{\alpha,\beta+\lambda,\gamma}}\\
&=&2\gamma^{\alpha/2}\,C_{\alpha,\beta,\gamma}\frac{K_{|\alpha|}(2\sqrt{\gamma(\beta+\lambda)})}{(\beta+\lambda)^{\alpha/2}}\\
&\sim&\sqrt{\pi} C_{\alpha,\beta,\gamma}\,\gamma^{(2\alpha-1)/4} \,\frac{\exp\{-2\sqrt{\gamma(\beta+\lambda)}\}}{(\beta+\lambda)^{(2\alpha+1)/4}}\,,\qquad \lambda\to\infty\,,
\end{eqnarray*}
as follows from~\eqref{Bessel}--\eqref{Besselass}. Observe $\Lambda_{\alpha,\beta,\gamma}(\lambda)\sim 2\sqrt{\gamma\lambda}$ as $\lambda\to\infty$
and thus $\sqrt{\pi}\,\Pi_{\alpha,\beta,\gamma}((x,\infty))\sim \Lambda_{\alpha,\beta,\gamma}(1/x)\sim 2\sqrt{\gamma/x}$ as $x\downarrow 0$ by  a Tauberian theorem~(see~\cite{b},~p.75) with associated Blumenthal-Getoor index 1/2.\halmos
\end{remark}
\begin{remark}\label{remCMYGasVGGGCconjecture}
Other possible extensions of the $VG^1$-class comprise a range of possible sample path
behaviour. In~\cite{MY08} the univariate $CGMY$-processes have been identified to be subordinated Brownian motions, and it is also known
that the associated subordinator is a $\GGC^1$-subordinator (see~\cite{JZ11}, their Example~8.2). As perceived in~\cite{CGMY02},
Blumenthal-Getoor indices of $CGMY$-processes exhaust the whole of the interval $(0,2)$. In particular,
for any given $q\in(0,1)$ there are $\GGC$-subordinators with Blumenthal-Getoor index $q$, the latter by Part~(a) of Proposition~\ref{propVGGCmoments} (see~\cite{LS10} for multivariate $CGMY$-models.)
 \halmos
\end{remark}
\begin{remark}\label{remnomomentsonlyFV} With~\eqref{thorinmeasure} being straightforwardly verified, $\TTT_\infty$ is a Thorin measure on $(0,\infty)$, where
\[\TTT_\infty(\rmd x)\,:=\,\eins_{(0,1/e)}(x) \frac{\rmd x}{x\log^{3}(1/x)}+  \eins_{(e,\infty)}(x) \frac{\rmd x}{\log^{2}(x)}\,.\]
As $\int_{x>1} \TTT_\infty(\rmd x)/x^{q}=\int_{0<x\le 1} \TTT_\infty(\rmd x)/x^{p}=\infty$ for $0\!<\!q\!<\!1$ and $p\!>\!0$, respectively, we get from Proposition~\ref{propGGCmoments}, that any associated $\GGC^1(a,\TTT_\infty)$-subordinator has infinite $p$-moments, with
Blumenthal-Getoor index equalling 1. By Proposition~\ref{propVGGCmoments}, also any $\VGGC^{d,1}(0,0,0,\Sigma,\TTT_\infty)$-process with $\det \Sigma>0$ has infinite $p$-moments, and
its Blumenthal-Getoor index equals 2.\halmos\end{remark}

\subsection{L\'evy Process Classes via Polar Decomposition}\label{subsecaltstelzerabreu}

Based on the polar decomposition of the L\'evy measure (see Lemma~\ref{lempolarrec}), P\'erez-Abreu and Stelzer~\cite{PS14} construct classes of self-decomposable distributions on cones
generating classes of L\'evy processes of finite variation ($FV^d$), surpassing subordinators, and including versions of Gamma and $GGC$ processes.\\[1mm]
\noindent{\bf Multivariate Gamma Process.}  We reproduce the model in~\cite{PS14} (their Section~3).
Recall $\mySS^d:=\{x\in\RR^d:\;\|x\|=1\}$, and let $\beta:\Sd\to(0,\infty)$ be a Borel function, and $\alpha$ a finite Borel measure on $\Sd$ such that
\begin{equation}\label{integragamma} \int_{\Sd}\,\log\frac{1+\beta(s)}{\beta(s)}\,\alpha(\rmd s)\quad<\quad\infty\,.
\end{equation}
We refer to a $d$-dimensional L\'evy process $X$ as a $\Gamma^d$-process with parameters $\alpha$ and $\beta$, written as $X\sim \Gamma^d_{L}(\alpha,\beta)$, whenever $X\sim FV^d(0,\Pi_X)$ with
\begin{equation}\label{LevymeasurePSGamma} \Pi_X\,=\,\int_{\Sd}\int_0^\infty
\dirac_{rs}\;e^{-\beta(s)r}\,\frac{\rmd r}r\;\alpha(\rmd s)\,.\end{equation}
Assuming $\alpha,\beta$ satisfying~\eqref{integragamma} it is shown in~\cite{PS14} that
the RHS in \eqref{LevymeasurePSGamma} defines a L\'evy measure (see their Proposition~3.3).  Much in the spirit of our Subsection~\ref{subsecmoments}, the $\Gamma^d_L$-class carries processes with infinite moments, amongst other things~(see~\cite{PS14}, their Examples~3.14, 3.15 and 3.16).\\[1mm]
\noindent{\bf Multivariate Gamma Subordinator.} Recall $\Sdplus:=[0,\infty)^d\cap \mySS^d$, and let $\beta:\Sdplus\to(0,\infty)$ be a Borel function, and $\alpha$ a finite Borel measure on $\Sdplus$ such that \eqref{integragamma} is satisfied, but with $\Sd$ replaced by $\Sdplus$.

We refer to a $d$-dimensional subordinator $T$ as a $\Gamma^d$-subordinator with parameters $\alpha$ and $\beta$, written as $T\sim \Gamma^d_{S}(\alpha,\beta)$, whenever, for all $\lambda\in[0,\infty)^d$,
\begin{equation}\label{defPSmultigamma}
-\log Ee^{-\skal\lambda {T(t)}}\,=\,
t\int_{\Sdplus}\,\log\frac{\beta(s)+\skal{\lambda}{s}}{\beta(s)}\,\alpha(\rmd s)\,.\end{equation}
As follows from the Frullani identity in~\eqref{laplacesubgamma}, the RHS in \eqref{defPSmultigamma} matches
\[\int_{\Sdplus}\,\log\frac{\beta(s)+\skal{\lambda}{s}}{\beta(s)}\,\alpha(\rmd s)\,=\,
\int_{\mySS^d_+}\int_0^\infty\left(1-e^{-r\skal\lambda s}\right)e^{-r\beta(s)}\frac{\rmd r}r\,\alpha(\rmd s)
\,,
\]
and, thus, $T\sim \Gamma^d_{S}(\alpha,\beta)$ holds if and only if $T\sim S^d(0,\Pi_T)$ with
$\Pi_T$ as in~\eqref{LevymeasurePSGamma}, but with $\Sd$ and $\Pi_X$ replaced by $\Sdplus$ and $\Pi_T$, respectively.

Plainly, a $\Gamma^d_S$-subordinator is a $\Gamma^d_L$-process with nondecreasing components: $S^d\cap \Gamma^d_L=\Gamma^d_S$.
In the univariate case, we have $\Gamma^1_{S}(\alpha\dirac_1 ,\beta)=\Gamma_{S}(\alpha,\beta(1))$.
The connection with our $GGC$-class is $\Gamma_S^d(\alpha,\beta)=GGC_S^d(0,\alpha\otimes_{p}\dirac_{\beta(\cdot)})$.\\[1mm]
The associated $V\Gamma^{d}_S$-class is thus a subclass of our $\VGGC^{d,d}$-class. Complementing this, it is possible to contrive the $V\Gamma^{d}_S$-class as a class of L\'evy processes associated with the {\em matrix-gamma-normal} class of~\cite{PS14} (see their Subsection~5.3.2).\\[1mm]
\noindent{\bf Multivariate Gamma Convolution Process.}~In~\cite{PS14} (see their Definition~4.4 and their Subsection~4.4) it is also shown how to characterise the associated class of multivariate generalised Gamma convolutions associated with cones. It is possible to introduce a $GGC^d_L$-class of $FV^d$-processes,
extending the polar decomposition from $[0,\infty)^d_*$ to $\RR_*^d$, as we illustrated in context of the inclusion $\Gamma_S^d\subseteq \Gamma_L^d$. In particular, we have $S^d\cap GGC^d_L=GGC_S^d$.

We conclude this subsection by establishing further inclusions in Figure~\ref{diagram}, postponing further investigations of this kind to Subsection~\ref{subsecaltsubordclass}.\\[1mm]
\noindent{\bf Variance Gamma ($\boldsymbol{VG}$) revisited.} Assume $Y\sim VG^d(b,\mu,\Sigma)$ such that $Y\sim \VGGC^{d,1}(0,\mu,\Sigma,\TTT)$ with $\TTT=b\dirac_b$ in~\eqref{defGVGd1}. If $d=1$ and $\Sigma\neq 0$ then~\eqref{GVGlevdensII} degenerates to, for $r>0$, $s\in \mySS_E^1=\{\pm 1\}$,
\[
g_Y(s,r)
\,=\,b\exp\Big\{r\big(s\mu-(2b\Sigma+\mu^2)^{1/2}|s|\big)/\Sigma\Big\}\Big/|s|\,,\]
and in~\eqref{LevymeasurePSGamma} we have \[Y\sim \Gamma_L^1\Big(b\,(\delta_{-1}+\delta_{1}),\pm 1\mapsto \big((2b\Sigma+\mu^2)^{1/2}\mp\mu\big)/\Sigma\Big)\,,\]
comparable to the representation of univariate $VG^1$-processes as a difference of independent Gamma subordinators in~\cite{MaCaCh}.

For $d>1$ and invertible $\Sigma$ we get from~\eqref{Besselass} and~\eqref{GVGlevdensII} that, for $s\in\mySS_E^d$, as $r\to\infty$,
\begin{eqnarray*}g_Y(s,r)
&=&\frac {2b}{(2\pi)^{d/2}}\frac{\exp\{r\skal{\Sigma^{-1}\mu} s\}}{(\det\Sigma)^{1/2}\|s\|^{d}_{\Sigma^{-1}}}\;\times
\widehat K_{d/2}\Big((2b\!+\!\|\mu\|^2_{\Sigma^{-1}})^{1/2} \; r\|s\|_{\Sigma^{-1}}\Big)\\
&\sim&\frac{b(2b\!+\!\|\mu\|^2_{\Sigma^{-1}})^{(d-1)/4}}{(2\pi)^{(d-1)/2}(\det\Sigma)^{1/2}}\times\frac{r^{(d-1)/2}}{\|s\|_{\Sigma^{-1}}^{(d+1)/2}}\\
&&\qquad\quad\times
\exp\Big\{r\big(\skal{\Sigma^{-1}\mu} s-(2b\!+\!\|\mu\|^2_{\Sigma^{-1}})^{1/2} \, \|s\|_{\Sigma^{-1}}\big)\Big\}\,.
\end{eqnarray*}
The asymptotic equivalence in the last display does not match~\eqref{LevymeasurePSGamma}, erasing the possibility of $Y$ being a $\Gamma^d_L$-process.

Within his $\VGGC^{d,1}$-class Grigelionis~(see~\cite{Gr07}, his Proposition~3) shows that any $\VGGC^{d,1}(0,\mu,\Sigma,\TTT)$-process
is self-decomposable, provided either $d=1$ or $d\ge 2$ and $\mu=0$. Assuming an invertible $\Sigma$ and imposing further moment conditions upon $\TTT$,
which are duly satisfied for finitely supported Thorin measures such as $\TTT=b\dirac_b$, he shows that his result is sharp: if
$\mu\neq 0$ and $d\ge 2$ then a $\VGGC^{d,1}(0,\mu,\Sigma,\TTT)$-process cannot be self-decomposable. Consequently, a $VG^d(b,\mu,\Sigma)$-process with $d\ge 2,\mu\neq 0$  and invertible $\Sigma$ cannot be self-decomposable, let alone be an element of the $\Gamma^d_L\subseteq\GGC_L^d$-classes.
Plainly, $\widehat K_{1/2}$ is completely monotone. By differentiating~\eqref{Bessel} under the integral sign we find that~$\widehat K_{d/2}''(0+)<0$. In particular, if 
$d\ge 2$, then $\widehat K_{d/2}$ is no longer completely monotone, and a $VG^{d}$-processes with $d\ge 2$, $\mu=0$ and invertible $\Sigma$ cannot be a $\GGC^{d}_L$-process.

It would be interesting to provide a detailed study regarding the $VGG^{d,k}$ ($k\in\{1,d\}$)-classes and the $\Gamma^d_L\subseteq GCC^d_L$-classes, but this is beyond the scope of our present paper.

\subsection{Exponential Moments and Esscher Transformation}\label{subsecEsscher}
We use the notation, for one hence all $t>0$,
\begin{equation}\label{defDDD} \DDD_{Y}=\{\lambda \in \RR^d:\, Ee^{\skal\lambda{Y(t)}}<\infty\}=\{\lambda\in\RR^d:\int_{\|y\|>1}\! e^{\skal{\lambda}{y}}\Pi_Y(\rmd y)<\infty\}.\end{equation}
$\DDD_{Y}$ is a convex subset of $\RR^d$, containing the origin (see~\cite{s}, p.~165), similarly we introduce $\DDD_T$.

Further, we need to introduce
\begin{equation}\label{defOlambda}
\OOO_\lambda:=\{x\in[0,\infty)^d_*:\,\|x\|^2>\skal\lambda x\}\,,\qquad \lambda\in\RR^d\,,
\end{equation}
and a transformation
\begin{equation}\SSS_\lambda(x)=\frac{\|x\|^2\!-\!\skal {\lambda} x}{\|x\|^2}\,x\,,\quad x\in\RR^d_*\,.\label{defS}\end{equation}
We provide conditions on the Thorin measure ensuring finiteness of exponential moments for the associated $GGC/VGG$-model (see Subsection~\ref{subsecproofEsscher} for a proof).
\begin{proposition}\label{propexpoGGCVGGC} Let $t>0$, $k\in\{1,d\}$, $\lambda\in\RR^d$, $T\sim GGC^d_S(a,\TTT)$ and $Y\sim \VGGC^{d,k}a,\mu,\Sigma,\TTT)$. Then:\\[1mm]
(a)~$\{0\}\cup ([0,\infty)_*^d\backslash\OOO_\lambda)$ is a convex and compact subset of $\RR^d$, and $\SSS_\lambda$ is a continuous function from $\OOO_{\lambda}$ into $[0,\infty)^d_*$.\\[1mm]
(b)~\quad$\lambda\in\DDD_T$\;\quad$\Leftrightarrow\qquad$ simultaneously,~$\TTT([0,\infty)_*^d\backslash\OOO_\lambda)=0$ and
\begin{equation}\label{eqGGCd2expmoments}
\int_{\OOO_\lambda}\log^-\|\SSS_\lambda(x)\|\,\TTT(\rmd x)=\int_{\OOO_\lambda} \log^-\frac{\|x\|^2\!-\!\skal\lambda x}{\|x\|}\,\TTT(\rmd x)\!<\!\infty\,.
\end{equation}
(c)~For $\lambda\in\DDD_T$ the image measure of the restriction $\TTT\big|\OOO_\lambda(\cdot):=\TTT(\OOO_\lambda\cap \cdot)$ under the mapping $\SSS_\lambda$,
denoted by $\TTT_\lambda:=(\TTT\big|\OOO_\lambda)\circ\SSS^{-1}_\lambda$, is a well-defined Thorin measure on $[0,\infty)^d_*$.\\[1mm]
(d)~Without restrictions on $(a,\mu,\Sigma,\TTT)$: $\lambda\in\DDD_Y\Leftrightarrow q_{\lambda,k}\in\DDD_T$, where
\begin{equation}\label{defqlambda}q_{\lambda,k}\,=\,\left\{\begin{array}{ll}\skal{\lambda}{\mu}+\frac 12\|\lambda\|^2_\Sigma\,,&\mbox{if }k=1\,,\\
\lambda\diamond\mu+\frac 12 \Sigma(\lambda\diamond\lambda)\,,&\mbox{if }k=d\,.
\end{array}\right.
\end{equation}
\end{proposition}
Assume that $Y\sim L^d(\gamma_Y,\Sigma_Y,\Pi_Y)$ is a L\'evy process with respect to an underlying stochastic basis $(\Omega,\FFF,\{\FFF_t\},P)$.

The {\em Esscher transform} on ${\cal F}_t$ with respect to $Y$ is given by
\begin{align}\label{defEsscher}
    \frac{\rmd Q^Y_{\lambda,t}}{\rmd P}
    \quad=\quad \frac{\exp\skal{\lambda}{Y(t)}}{E_P\exp\skal\lambda{Y(t)}} \, ,
    \qquad t\ge 0\,,\;\lambda\in\DDD_Y\,.
\end{align}
For $t\ge 0$ and $\lambda\in\DDD_Y$ in~\eqref{defDDD} it is well-known that $Q^Y_{\lambda,t}:\FFF_t\to [0,1]$ defines a probability measure, equivalent to $P:\FFF_t\to[0,1]$.
Besides this, $\{Y(s):\,0\le s\le t\}$ remains a L\'evy process under the new measure $Q^Y_{\lambda,t}$.

Next we show that both $\VGGC$-classes are {\em invariant} under Esscher transformations~(for a proof see Subsection~\ref{subsecproofEsscher}) We provide more specific examples throughout the remaining part of the paper, more specifically see Theorem~\ref{theoVMG}, Subsection~\ref{subsecaltsubordclass} and Section~\ref{secapplic}.
\begin{theorem}\label{theoEsscherVGGC} Let $t\!\ge\! 0$, $k\!\in\!\{1,d\}$. If $Y\sim \VGGC^{d,k}(a,\mu,\Sigma,\TTT)$ with $\lambda\in\DDD_Y$
then $q:=q_{\lambda,k}\in\DDD_T$ and
$\{Y(s):0\!\le\! s\!\le\! t\}|Q^Y_{\lambda,t}\sim\VGGC^{d,k}(a,\mu+\Sigma\lambda,\Sigma,\TTT_q)$ with $q_{\lambda,k}\in\RR^k$ and $\TTT_q$ as in~Proposition~\ref{propexpoGGCVGGC}.
\end{theorem}
\subsection{$\boldsymbol{V\MMM\Gamma}^d$-Class}
\label{subsecVGamma}
In this subsection we restrict ourselves to {\em finitely supported} Thorin measures and consider a corresponding subclass of $\VGGC^{d,d}$.\\[1mm]
\noindent{\bf $\boldsymbol{\MMM\Gamma^d}$-Subordinator.} Let
$n\in\NN=\{1,2,\dots\}$, $b_*=(b_1,\dots,b_n)'\in(0,\infty)^n$, $M=(m_{kl})_{1\le k\le d,1\le l\le n}\in\RR^{d\times n}$ having columns $M_1,\dots,M_n\in[0,\infty)^d_*$ and let $G_1\sim \Gamma_S(b_1),\dots,G_n\sim\Gamma_S(b_n)$ be independent standard Gamma processes.

We call a $d$-dimensional subordinator $T$ an {\em $\MMM\Gamma^d$-subordinator} with parameters $n$, $b_*$, $M$, briefly $T\sim \MMM\Gamma_{S}^d(b_*,M)$, provided
\begin{equation}\label{defmultigamma}T\eqd M(G_1,\dots,G_n)'\,=\,\sum_{l=1}^n G_l M_l\,.\end{equation}
Next, we show that $\MMM\Gamma^d$-subordinators are $GGC^d$-subor\-dinators, but having zero drift $a=0$ and finitely supported Thorin measure:
\begin{lemma}\label{lemsubgamma} Let $T\sim \MMM\Gamma^d_{S}(b_*,M)$ with $n=\dim b_*$. Then~$T\sim S^d(0,\Pi_T)=\GGC^d_S(0,\TTT_T)$, where, simultaneously,
\begin{eqnarray}
\label{Mgammalevmeas}
\Pi_{T}&=&\sum_{l=1}^n b_l \int_{(0,\infty)}\dirac_{r M_l}\;\exp\{-b_l r\}\rmd r\big/r\,,\\
\TTT_{T}&=&\sum_{l=1}^n b_l\,\dirac_{b_lM_l/\|M_l\|^2}\,,\label{MgammaThormeas}\\
\DDD_T&=&\bigcap_{l=1}^n\{\lambda\in\RR^d:\,\skal {M_l}\lambda<b_l\}\label{MgammaDDDT}\end{eqnarray}
and, for $t\ge 0$, $\lambda\in\DDD_T$,
\begin{equation}\label{Mgammalapl}-\log E e^{\skal{\lambda}{T(t)}}\,=\,t\sum_{l=1}^n b_l\,\log\big\{(b_l-\skal {M_l}\lambda )\big/b_l\big\}\,.\end{equation}
\end{lemma}
\noindent{\it Proof.} By~\eqref{LaplaceGGCd},~\eqref{Mgammalapl} follows straightforwardly from~\eqref{MgammaThormeas}.
Part~(a) of~Proposition~\eqref{propdecomp} is applicable to~\eqref{defmultigamma} giving $\Pi_T=\sum_{l=1}^n \Pi_{G_lM_l}$, with a similar superposition and intersection in place for $\TTT_T$ and $\DDD_T$, respectively. It remains to verify the formulae in~\eqref{Mgammalevmeas}--\eqref{MgammaDDDT}, but with $n\!=\!1$, $M\!=\!M_1$, $b\!=\!b_1$ and $G\!=\!G_1\sim \Gamma_S(b)$.

With $\Pi:=b\int_{(0,\infty)}\dirac_{r M} e^{-b r}\;\rmd r/r$ and $\lambda\in\RR^d$ with $\skal\lambda M<b$ calculate
\begin{equation}
\int_{\RR^d_*} (1\!-\!e^{\skal{\lambda}x})\,\Pi(\rmd x)\,
=\,b\int_{0}^\infty \left(1-e^{r\skal\lambda M}\right)\,
\;\;e^{-b r}\;\rmd r\big/ r\,.\label{lapexpPiGxi}
\end{equation}
On the other hand, $E\exp\skal\lambda {G(t) M}=
E\exp\{\skal\lambda M G(t)\}$
in which we can substitute the characteristic exponents of $G$
using~\eqref{laplacesubgamma}. By~\eqref{lapexpPiGxi}, the resulting expressions match those in~\eqref{Mgammalevmeas}--~\eqref{MgammaThormeas}.\halmos\\[1mm]
For the remaining part we review some properties of the Gamma distribution.
\begin{lemma}\label{lemgamma} Let $c,\alpha,\beta,\alpha_1,\dots,\alpha_n,\beta_1,\dots,\beta_n>0$. Let $Z\sim \Gamma(\alpha,\beta)$. Let
$Z_1,\dots,Z_n$ be independent with $Z_k\sim \Gamma(\alpha_k,\beta_k)$
for $1\le k\le n$. Then we have $c Z\sim\Gamma(\alpha,\beta/c)$ as well as the equivalence `(i)$\Leftrightarrow$(ii)', where: (i)~$\sum_{k=1}^nZ_k\,\sim\, \Gamma(a,b)$ for some $a,b>0$; (ii) \, $\beta_1=\dots=\beta_n$. If (i) or (ii) is satisfied then $b=\beta_1$ and $a=\sum_{k=1}^n \alpha_k$.
\end{lemma}
\noindent{\it Proof.}~Note $a\dirac_b= \sum_{k=1}^n \alpha_k\dirac_{\beta_k}$ holds if and only if both $b=\beta_1={\dots}=\beta_n$ and $a=\sum_{k=1}^n \alpha_k$, and the equivalence~'(i)$\Leftrightarrow$(ii)' follows from the Thorin representation of the Gamma distribution.\halmos\\[1mm]
We introduced the $\Gamma^d_{S}(\alpha,\beta)$ in~\eqref{defPSmultigamma}, and its connection with the $GGC$-class was  $\Gamma_S^d(\alpha,\beta)=GGC_S^d(0,\alpha\otimes_{p}\dirac_{\beta(\cdot)})$ with the identity $\Gamma^1_{S}(\alpha\dirac_1 ,\beta)=\Gamma_{S}(\alpha,\beta(1))$ in the univariate case~$(d=1)$.

As we clarify next, a given $\MMM\Gamma^d_S$-subordinator does not need to have Gamma marginals nor does it need to be $\Gamma^d_S$-subordinator.
\begin{lemma}\label{lemsubgammaIII} Let~$T=(T_1,\dots, T_d)'\sim \MMM\Gamma^d_{S}(b_*,M)$ with $n=\dim b_*$.\\[2mm]
(a)~Then (i) $\Leftrightarrow$ (ii), where (i) \, $T_k\sim \Gamma_S(p_k,q_k)$
for some $p_k,q_k\in(0,\infty)$;\\[1mm]
(ii) \, there exists $1\!\le\!l_0\!\le\!n$ with $m_{k,l_0}>0$ such that $b_lm_{k,l_0}=b_{l_0}m_{k,l}$ for all $1\!\le\! l\!\le\! n$ with $m_{k,l}\neq 0$.\\[2mm]
(b)~(i') $\Leftrightarrow$ (ii'), where~(i') \, $T\sim \Gamma_{S}^d(\alpha,\beta)$ for some $\alpha,\beta$;\\
(ii') \, for all $1\!\le\! k,l\!\le\! n$, the following implication holds
\begin{equation}\label{Mgammaisgamma} \|M_l\| M_k=\|M_k\|M_l\quad\Rightarrow\quad \|M_l\|b_k=\|M_k\|b_l\,.
\end{equation}
In addition, if one of (i) or (ii) holds then we have $q_k=b_{l_0}/m_{k,l_0}$ and $p_k=\sum_{m_{k,l}\neq 0}b_l$. Also,
if one of (i') or (ii') is satisfied then $\alpha=\sum_{l=1}^n b_l \dirac_{M_l/\|M_l\|}$ and $\beta(M_l/\|M_l\|)=b_l/\|M_l\|$ $(1\!\le\! l\!\le\! n)$.
\end{lemma}
\noindent{\it Proof.}~(a)~follows from Lemma~\ref{lemgamma} as we can decompose the $k$th component of $T$ in~\eqref{defmultigamma} into a sum of $n$ univariate Gamma subordinators.\\[1mm]
(b)~`(ii')$\Rightarrow$(i')': For $S\sim\Gamma^d_S(\alpha,\beta)$ and $x\in[0,\infty)^d$ with Euclidean norm $\|x\|_E^2=\skal{x}{x}=1$, introduce a univariate subordinator $S^{x}$ by
\begin{equation}\label{defSx}S^{x}(t)\,:=\,\sum_{0<s\le t}\,\eins_{\{\alpha x:\,\alpha>0\}}(\Delta S(s))\,\skal{x}{\Delta S(s)}\,,\qquad t\ge 0\,.\end{equation}
To determine the L\'evy measure $\Pi^x$ of $S^x$, let $A\subseteq (0,\infty)$ be a Borel set and note
\begin{eqnarray*}
\Pi^{x}(A)&=&E[\#\{0\le t\le 1: \Delta S(t)\in\{\alpha x:\,\alpha>0\}~\mbox{and}~\skal{x}{\Delta S(t)}\in A \}]\\
&=&\Pi_S(\{\tau\in[0,\infty)^d_*:\,\tau \in\{\alpha x:\,\alpha>0\}~\mbox{and}~\skal{x}{\tau}\in A \})\,.
\end{eqnarray*}
As $\eins_{\{\alpha x:\,\alpha>0\}}(rs)=\eins_{\{\alpha x:\,\alpha>0\}}(s)=\eins_{\{x/\|x\|\}} (s)$ for $r>0,s\in\mySS_d^+$,
we get from \eqref{LevymeasurePSGamma} that the RHS in the last display matches
\begin{eqnarray*}
\Pi^{x}(A)&=&\int_{\mySS_+^d}\int_0^\infty
\eins_A(r \skal xs)\;
\eins_{\{x/\|x\|\}}(s)\,\;e^{-\beta(s)r}\,\frac{\rmd r}r\;\alpha(\rmd s)\\
&=&\alpha(\{x/\|x\|\}) \int_0^\infty
\eins_A(r/\|x\|)\,\;e^{-\beta(x/\|x\|)r}\,\rmd r\big/r\,.
\end{eqnarray*}
Substituting $r=r'\|x\|$ on the RHS of the last display, it follows from~\eqref{laplacesubgamma} that either $S^x=0$ or $S^{x}\sim\Gamma_S(\alpha(\{x/\|x\|\}),\|x\|\beta(x/\|x\|))$.

Thus prepared, let $T\sim \MMM\Gamma^d_{S}(b_*,M)$ with $T\eqd \sum_{l} G_lM_l$ as in~\eqref{defmultigamma} for independent standard Gamma subordinators $G_1,\dots,G_n$. Let $T^x$ be defined as in~\eqref{defSx}, but with $S$ replaced by $T$.
In particular,
$$T^{M_k/\|M_k\|_E}\eqd\sum_{\|M_l\|M_k=\|M_k\|M_l} G_l \skal{M_k}{M_l}/\|M_k\|_E\,,\quad 1\le k\le n\,.$$
In addition, suppose $T\sim \Gamma_S^d(\alpha,\beta)$. Then $T^{M_k/\|M_k\|_E}$ must either be degenerate or a univariate Gamma subordinator. Consequently, by Lemma~\ref{lemgamma}, we must have $b_l\|M_k\|_E^2=
b_k\skal{M_k}{M_l}$ for $1\le l\le n$ with $\|M_l\|M_k=\|M_k\|M_l$. The latter is equivalent to (ii'),
completing the proof of `(ii')$\Rightarrow$(i')'. The proof of `(i')$\Rightarrow$(ii')' is analogous.\halmos\\[1mm]
\noindent{\bf Variance--$\boldsymbol{\MMM\Gamma}$ $(\boldsymbol{V\MMM\Gamma}^d$).} With parameters $b_*=(b_1,\dots,b_n)'\in(0,\infty)^n$ and $M\in\RR^{d\times n}$ as set for an $\MMM\Gamma^d$-subordinator, in addition, take $\mu=(\mu_1,\dots,\mu_d)'\in\RR^d$ and a {\em diagonal} matrix $\Sigma=\diag(\Sigma_{11},\dots,\Sigma_{dd})$ with nonnegative entries.

Whenever $Y\eqd B\circ_{d}T$, with $B,T$ being independent and $B\sim BM^d(\mu,\Sigma)$ being Brownian motion, while  $T\sim \MMM\Gamma_{S}^d(b_*,M)$, we call $Y$ a {\em Variance $\MMM\Gamma$ $($VM$\Gamma^d$$)$-process}, written in the following as
\begin{equation}\label{defGenSVG}Y\,\sim\, V\MMM\Gamma^d(b_*,M,\mu,\Sigma):=BM^{d}(\mu,\Sigma)\circ_{d}\MMM\Gamma^d_{S}(b_*,M)\,.\end{equation}
For a generic case, where $\det\Sigma\neq 0$, we give formulae for the canonical L\'evy measure $\Pi_Y$. To each column $M_l$ we associate both a dimension $1\!\le\!d_l\!\le\!d$ by
\[d_l\,:=\,\#\{1\le k\le d: m_{kl}>0\}\,,\qquad 1\le l\le n\,,\]
and a $\sigma$-finite Borel measure $\ell^*_l:=\bigotimes_{k=1}^d \ell^*_{l,k}$ on $\RR^d$ as a product measure with the following factors
\begin{equation}\label{defmukl}
\ell^*_{l,k}\,:=\,\eins_{(0,\infty)}(m_{kl})\ell+\eins_{\{0\}}(m_{kl})\dirac_{0}\,,\quad 1\!\le\! k\!\le\! d,\;1\!\le\! l\!\le\! n\,.\end{equation}
For $1\le l\le n$, we set
\begin{eqnarray}\nonumber
\beta_l&:=&2b_l\!+\!\sum_{m_{k,l}\neq 0}m_{kl}\mu^2_{k}\big/\Sigma_{kk}\,=\,2b_l\!+\!\skal{\mu\diamond M_{l}}{\Sigma^{-1}\mu}\,,\nonumber\\
\alpha_l&:=&\big(2^{(2-d_l)/2}\;\pi^{-d_l/2}\,b_l\;\beta_l^{d_l/4}\big)\;\bigg/\prod_{m_{kl}\neq 0}\Sigma_{kk}^{1/2}m^{1/2}_{kl}\,. \label{constPiYSVG}
\end{eqnarray}
The next theorem gives formulae for the L\'evy measure and Laplace exponent of $Y$, which has finite variation (recall~\eqref{PiFVintegrab}) and is invariant in form under Esscher tranformations.
\begin{theorem}\label{theoVMG}Assume $Y\sim V\MMM\Gamma^d(b_*,M,\mu,\Sigma)$ with $n=\dim b_*$. Then:\\[2mm]
(a)~We have $Y\sim\VGGC^{d,d}\big(0,\mu,\Sigma,\TTT\big)$ with $\TTT=\sum_{l=1}^n b_l\dirac_{b_lM_l/\|M_l\|^2}$.\\[1mm]
(b)~Always, $Y\sim FV^d(0,\Pi_Y)$. When, in addition, $\det(\Sigma)>0$, then for all Borel sets $A\subseteq\RR^d_*$,
\begin{eqnarray*}&&\Pi_{Y}(A)\,=\,\\
&&\sum_{l=1}^n\alpha_l \int_A \frac{K_{d_l/2}\left(\sqrt{\beta_l\sum_{m_{kl}\neq 0}y_k^2/m_{kl}}\right)}{\left(\sum_{m_{kl}\neq 0}y_k^2/(\Sigma_{kk} m_{kl})\right)^{d_l/4}}
\exp\bigg\{\sum_{m_{kl}\neq 0}\mu_k y_k/\Sigma_{kk}\bigg\}\;\ell^*_l(\rmd y)\,.\nonumber
\end{eqnarray*}
(c)~We have\[\DDD_Y=\big\{ \lambda\in\mathbb{R}^d:  \langle\mu \diamond M_l,\lambda\rangle + \frac{1}{2}\|\lambda\|^2_{\Sigma\diamond M_l} < b_l,\quad 1\le l \le n\big\}\,,\]
and, for $t\!\ge\!0$ and $\lambda\in\DDD_Y$,
\begin{equation}\label{LaplaceVMG}
-\log Ee^{\skal \lambda {Y(t)}}
=t\sum_{l=1}^n b_l\log\Big\{\big(b_l\!-\!\!\skal{\mu\diamond M_l}\lambda\!-\!\frac 12\|\lambda\|^2_{\Sigma \diamond M_l}\big)/b_l\Big\}\,,
\end{equation}
and
\[\{Y(s):0\!\le\! s\!\le\! t\}|Q^Y_{\lambda,t}\quad\sim\quad V\MMM\Gamma^d(b_*,M_\lambda,\mu_\lambda,\Sigma)\,.\]
Here $\mu_\lambda=\mu+\Sigma\lambda$, and $M_\lambda\in[0,\infty)^{d\times n}$ has the following columns $M^{\lambda}_1\dots,M^\lambda_n$:
\begin{equation}M^\lambda_l= \frac{b_l}{b_l - \langle\mu \diamond M_l,\lambda\rangle - \frac{1}{2}\|\lambda\|^2_{\Sigma\diamond M_l}} \, M_l\, , \qquad 1 \le l\le n\, .\label{defMlambda}
\end{equation}
\end{theorem}
\noindent{\it Proof.}~(a)~follows from Lemma~\ref{lemsubgamma}~and~\eqref{defGenSVG}. $\MMM\Gamma^d$-subordinators $T$ have zero drift. Also, $\int_{0<\|x\|\le 1}\|x\|^{1/2}\Pi_T(\rmd x)$ is finite for $\MMM\Gamma^d$-subordinators $T$. Thus, $Y\sim FV(0,\Pi_Y)$~by Part~(c) of Lemma~\ref{lemunifunimultisub}. (A stronger result follows from Part~(a) in Proposition~\ref{propVGGCmoments}.) In view of Part~(a), the remaining parts~(b)-(c) follow from Theorems~\ref{theoGVGmulti}-\ref{theoEsscherVGGC} as well as Proposition~\ref{propexpoGGCVGGC}.\halmos
\begin{remark}\label{remWang} If $Y\sim V\MMM\Gamma^d(b_*,M,\mu,\Sigma)$ with $n=\dim b_*$ then
 \begin{equation}\label{superposVG}Y\eqd\sum_{l=1}^n Y_l\,,\end{equation}
as an implication of~\eqref{LaplaceVMG}. Here $Y_1,\dots,Y_n$ are independent processes with $Y_l\sim VG^d(b_l,\mu\diamond M_l,\Sigma\diamond M_l)$ for $1\le l\le n$. It is, thus, possible to construct $V\MMM\Gamma^d$-processes by superimposing independent Madan-Seneta $VG^d$-processes. Wang~\cite{Wa09} comes to similar conclusions, and constructs multivariate L\'evy processes with $VG^1$-components by superimposing suitable $VG^d$-processes, just as in the right hand-side of~\eqref{superposVG}. \halmos
\end{remark}
\noindent{\bf Transition Densities.} For a subclass of $\MMM\Gamma^d$-class it is possible
to obtain formulae for transition densities, as we illustrate next.

Let $b_*=(b_1,\dots,b_{d+1})'\in(0,\infty)^{d+1}$ and $M\in[0,\infty)^{d\times (d+1)}$
such that, simultaneously,
\begin{eqnarray}M\,=\,(m_{kl})_{1\le k\le d, \,1\le l\le d+1}&=&\big(\diag(m_{11},\dots,m_{dd}),M_{d+1}\big)\quad\mbox{and}\nonumber\\
\prod_{k=1}^d m_{kk}\; m_{k(d+1)}&\neq&0\,.\label{gensubclasssemeraroI}
\end{eqnarray}
With $t>0$ define
\begin{eqnarray}\label{defCtMb}
C^*_t&:=&C^*_t(b_*,M)\,:=\,\left\{b_{d+1}^{t b_{d+1}}\big/\Gamma(tb_{d+1})\right\}\prod_{k=1}^d b_k^{t b_k}\big/\big(\Gamma(t b_k)m_{k,k}^{t b_k}\big)\,,\\
\beta^*&:=&\beta^*(b_*,M)\,:=\,-b_{d+1}+\sum_{k=1}^db_km_{k,d+1}\big/m_{kk}\label{defbetabM}
\,.
\end{eqnarray}
The proof of the next result follows from a similar analysis as in Section~48.3.1 in~\cite{KBJ}. (We are unable to provide substantial simplification of the integral in \eqref{densmultgamma} which occurs by integrating the joint density of $d\!+\!1$ independent Gamma random variables. However, using the results in~\cite{MM91}, it is possible to expand the integral in terms of Lauricella functions.)
\begin{lemma}\label{lemtransitiondensmultgam}
 Let $t>0$ and $T\sim \MMM\Gamma^d_S(b_*,M)$ with $M$ satisfying~\eqref{gensubclasssemeraroI}.
Then $T(t)$ admits a Lebesgue density $f_T$: for $\tau=
(\tau_1,\dots,\tau_{d})'\in\RR^{d}$,
\begin{eqnarray}
\lefteqn{f_{T(t)}(\tau)
\,=\,C_t^*\; 1_{(0,\infty)^d}(\tau) \exp\big\{-\sum_{k=1}^db_k\tau _k\big/m_{kk}\big\}\times}\nonumber&&\\
&&\int_{0}^{\min_{1\le k\le d} \tau_k/m_{k,d\!+\!1}}\,
e^{\beta^* s}\;s^{t b_{d+1}\!-\!1}\prod_{k=1}^d (\tau_k\!-\!m_{k(d+1)}s)^{tb_k\!-\!1}\;\rmd s\,.\label{densmultgamma}
\end{eqnarray}
\end{lemma}
Next, we state such formulae for the associated $V\MMM\Gamma^d$-model. Let $\mu=(\mu_1,\dots,\mu_d)'$ and $\Sigma=\diag(\Sigma_{11},\dots,\Sigma_{dd})$ be the parameters of the underlying Brownian motion. To ensure the existence of a Lebesgue transition density for the $V\MMM\Gamma^d$-process, we need to make the additional assumption that $\Sigma$ is invertible, $i.e.$
all $\Sigma_{kk}>0$.

With the help of~\eqref{defCtMb} and~\eqref{defbetabM} define
\begin{eqnarray*}
a_k:=1\big/(2m_{k(d+1)}\Sigma_{kk})\,,&&
\widehat a_k:=m_{k(d+1)}\;\big((b_k/m_{kk})\!+\!(\mu_k^2/(2\Sigma_{kk}))\,,\\
c_k:=2 b_k+m_{kk}\mu_k^2/\Sigma_{kk}\,,&&
\widehat c_k:=\sqrt{c_k/(\Sigma_{kk}m_{kk})}\,.
\end{eqnarray*}
Further, for $t>0$, we set
\begin{eqnarray*}
C_t&:=&C^*_t(b_*,M)\;(2^d\pi^d \det\Sigma)^{-1/2}\prod_{k=1}^d m_{k(d+1)}^{tb_k-\frac 12}\nonumber\\
&=&\left\{b_{d+1}^{t b_{d+1}}\big/\big(2^{d/2}\pi^{d/2}\;\Gamma(tb_{d+1})\big)\right\}\;\;
\prod_{k=1}^db_k^{t b_k}m_{k(d+1)}^{tb_k-\frac 12}\big/\big(\Sigma_{kk}^{1/2}\Gamma(t b_k)\,m_{kk}^{t b_k}\big)\,,\\
c_{d+1}&:=&2 b_{d+1}+\sum_{k=1}^dm_{k(d+1)}\mu_k^2\big/\Sigma_{kk}\,,\\
D_t&:=&2\pi^{-d}\;c_{d+1}^{(d-2 b_{d+1}t)/4}\;\prod_{k=1}^{d+1}
\big(b_k^{b_k t}\big/\Gamma(b_k t)\big)\;\times\\
&&{}\times\prod_{k=1}^d\Big\{c_k^{(1-2 b_k t)/4}\;\Sigma_{kk}^{-(3+2b_kt)/4}\;m_{kk}^{-(1+2 b_k t)/4}\;m_{k(d+1)}^{-1/2}\Big\}\,.\nonumber
\end{eqnarray*}

\begin{theorem}\label{theotransitiondenSemVG} Assume $Y\sim
V\MMM\Gamma^d(b_*,M,\mu,\Sigma)$ with $\det\Sigma\neq 0$ and~\eqref{gensubclasssemeraroI}.
Then the law of $Y(t)$ admits a Lebesgue density $f_{Y(t)}$ for $t\!>\!0$:
\begin{eqnarray}
\lefteqn{f_{Y(t)}(y)\,=\,C_t \exp\big\{\sum_{k=1}^d \mu_k y_k\big/\Sigma_{kk}\big\}\times}&&\label{GSVGdensI}\\
&&\hspace*{-2em}\int_0^\infty\!\!\;
e^{\beta^* s}s^{-\frac {d+2}2-t \sum_{k=1}^{d+1}b_{k} }\,\prod_{k=1}^d\int_{0}^1
\exp\!\!\left\{\!-
\frac{a_kuy^2_k}s-\frac{\widehat a_k s}u\!\right\}\frac{(1\!-\!u)^{t b_k-1}}{u^{t b_k+\frac 12}}\rmd u\;\rmd s\nonumber\\
\nonumber
\end{eqnarray}
\begin{eqnarray}
&&\,=\,D_t\exp\big\{\sum_{k=1}^d\mu_k y_k\big/\Sigma_{kk}\big\}\times\!\!\label{GSVGdensII}\\
\nonumber&&\hspace*{-2em}\int_{\RR^d}
\frac{K_{|2 b_{d+1}t-d|/2}
\big(
\sqrt{c_{d+1}\sum_{k=1}^dz_k^2\big/(\Sigma_{kk}m_{k(d+1)})}
\big)}{\big(\sum_{k=1}^dz_k^2\big/(\Sigma_{kk}m_{k(d+1)})\big)^{(d-2 b_{d+1} t)/4}}\prod_{k=1}^d
\frac{K_{|2b_k t-1|/2}\left(\widehat c_k|y_k-z_k|\right)\,\rmd z_k}{|y_k-z_k|^{(1-2b_k t)/2}}\nonumber
\end{eqnarray}
for $y=
(y_1,\dots,y_{d})'\in\RR^{d}$.
\end{theorem}
\noindent{\it Proof.} Since we have $Y\eqd B\circ_{d} T$ for independent $B\sim BM^d(\mu,\Sigma)$ and  $T\sim \MMM\Gamma^d(b_*M)$,~\eqref{GSVGdensI} follows from Lemma~\ref{lemtransitiondensmultgam} by conditioning $Y(t)=B(T(t))$ on  values of $T(t)$
(see Part~(b) of~Lemma~\ref{lemunifunimultisub}). In view of Remark~\ref{remWang}, we may write $Y\eqd Y_1+Y_2$ for independent
processes $Y_1$ and $Y_2$. Here the $d$-dimensional process $Y_1$ has {\em independent} components with the
$k$th component being a $VG^1(b_k,\mu_km_{kk},\Sigma_k m_{kk})$-process ($1\!\le\! k\!\le\! d$). Further, $Y_2$ is a $VG^d(b_{d+1},\mu\diamond M_{d+1},\Sigma\diamond M_{d+1})$-process. The formula in~\eqref{GSVGdensII} follows from~\eqref{VGlaw} by convolution.\halmos
\subsection{Subclasses of $GGC$-Subordinators}\label{subsecaltsubordclass}
\begin{center}
\begin{figure}[hpb!]
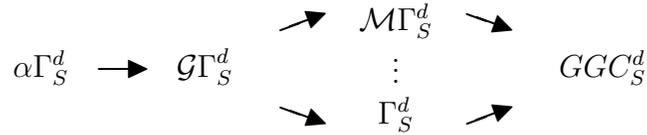

\[\begin{array}{cccccc}
{\alpha\Gamma^d_S}&
\BBarrow{000}&
{\GGG\Gamma}^d_S&
\begin{array}{ccc}
\BBarrow{20}&\MMM\Gamma^d_S&\BBarrow{340}\\&\vdots&\\
\BBarrow{340}&\Gamma^d_S&\BBarrow{20}
\end{array}&
\GGC^{d}_S
\end{array}
\]
\caption{\it An arrow points in the direction of generalisations of different subordinator classes, as described in the text.
${\cdots}$~indicates inclusion in special cases.}\label{diagramsubord}
\end{figure}
\end{center}

In this subsection we review subordinator classes as they occur in the literature and relate them to our formulations. The $\GGC^d$, $\Gamma^d$,  and $\MMM\Gamma^d$-subordinator classes were
introduced in Subsections~\ref{subsecVGGC} and~\ref{subsecVGamma}, respectively, and one of the connections was  $\Gamma_S^d(\alpha,\beta)=GGC_S^d(0,\alpha\otimes_{p}\dirac_{\beta(\cdot)})$. As defined in~\eqref{defmultigamma}, in Lemma~\ref{lemsubgamma} the $\MMM\Gamma^d$-class was identified
to be the subclass of $\GGC^{d}$-subordinators with drift $a=0$, having finitely supported Thorin measures $\TTT$.
As we clarified in Part~(b) of Lemma~\ref{lemsubgammaIII} a given $\MMM\Gamma^d$-subordinator does not need to be a $\Gamma^d$-subordinator. Two other classes, such as the ones introduced by
Semeraro~\cite{Se08} and Guillaume~\cite{Gu13}, are related to them as shown in Figure~\ref{diagramsubord}.  (Compare Figure~\ref{diagramsubord} with Figure~\ref{diagram}.) In the univariate case, where $d=1$, note that $\alpha\Gamma^1_S=\Gamma^1_S=\Gamma_S$.
\noindent{\bf  P\'erez-Abreu and Stelzer's $\boldsymbol{\Gamma^d}$-Subordinator.} The $\Gamma^d_S$-class of subordinators based on~\cite{PS14} was defined in~\eqref{defPSmultigamma}. In the univariate case, we have observed $\Gamma^1_{S}(\alpha\dirac_1 ,\beta)=\Gamma_{S}(\alpha,\beta(1))$.
The connection with our $GGC$-class was $\Gamma_S^d(\alpha,\beta)=GGC_S^d(0,\alpha\otimes_{p}\dirac_{\beta(\cdot)})$.

Let $T\sim \Gamma^d_{S}(\alpha,\beta)$, $\lambda\in\RR^d$, $q=q_\lambda\in\DDD_T$ as in~\eqref{defqlambda}. We get from Part~(a) of Proposition~\ref{propexpoGGCVGGC} that,  simultaneously, $\alpha\{\beta(\cdot)\le \langle q,\cdot\rangle\})=0$ and~\eqref{integragamma} holds
 with $\beta$ replaced by $\beta_\lambda(\cdot):=\beta(\cdot)- \langle q,\cdot\rangle$. In Part~(c) of Proposition~\ref{propexpoGGCVGGC}, observe that
$(\alpha\otimes_p \dirac_{\beta(\cdot)})_\lambda=\alpha\otimes_p \dirac_{\beta_\lambda}$. Consequently, the associated $V\Gamma_S^d$-class of subordinated Brownian motions is {\em closed under the Esscher} transform in the interpretation of Theorem~\ref{theoEsscherVGGC}.\\[1mm]
{\bf Semeraro's $\boldsymbol{\alpha}$-Subordinator.} Semeraro~\cite{Se08} introduced another approach to multivariate Gamma subordinators~(also see~\cite{LS10,LS10a,LS10b}). The parameters of this model are as follows: let $a,b\in(0,\infty),\alpha_*=(\alpha_1,\dots,\alpha_d)'\in(0,\infty)^d$ such that, simultaneously, $b>a\alpha_k$ for all $1\!\le\! k\!\le\! d$. Let
$S_1,\dots, S_{d+1}$ be independent
such that

\[
S_k\,\sim\,\Gamma_S\Big(\frac{b}{\alpha_k}-a,\,\frac b{\alpha_k}\Big)\,,\;\;1\le k\le d\,,\;\quad\quad
S_{d+1}\,\sim\,\Gamma_S(a,b)\,.\]
We refer to $T$ as an {\em $\alpha$-subordinator}, in brief $T\sim \alpha\Gamma^d_S(a,b,\alpha_*)$, provided
$T\eqd (T_1,\dots,T_d)'$ with \begin{equation}\label{defsemeraro}T_k:= S_k+\alpha_k S_{d+1}\,.\end{equation}

Any $\alpha \Gamma^d$-subordinator $T$ admits standard Gamma mar\-ginal distributions: $T_k\sim\Gamma_S(b/\alpha_k)$. As a result, the associated $V\alpha\Gamma^d_S$-processes, called $\alpha VG$ in \cite{Se08} have $VG^1$-marginal distributions.

We give an alternative representation of $T$ in~\eqref{defsemeraro}. Introduce para\-meters $b_*=(b_1,\dots,b_{d+1})'\in(0,\infty)^{d+1}$ and independent standard Gamma subordinators
$G_1,\dots,G_{d+1}$, $G_k\sim \Gamma_S(b_k)$ for $1\!\le\! k\!\le\! d\!+\!1$, by setting
\[b_k\,:=\,\frac{b}{\alpha_k}-a\,,\quad 1\le k\le d\,,\quad\quad
b_{d+1}:=a\,,\]
and, with $S_1,\dots,S_{d+1}$ as above,
\[G_k:=\frac b{b\!-\!a\alpha_k} S_k\,,\quad 1\le k\le d\,,\quad\quad G_{d+1}:=\frac ba S_{d+1}\,.\]
For $T$ in~\eqref{defsemeraro} we conclude that $T\sim \MMM\Gamma^d_S(b_*,M(a,b,\alpha_*))$, where in our notation
\begin{equation}\label{semeraroM}M(a,b,\alpha_*)\,:=\,\Big(\diag(b-\!a\alpha_1,\dots,b\!-\!a\alpha_d)\,,\;
a\alpha_*\Big)\big/ b\quad\in[0,\infty)^{d\times (d\!+\!1)}_*\,.\end{equation}

We show that the $V\alpha\Gamma^d_S$ process is not
closed under Esscher transform by considering the following bivariate example. In Part~(c) of Theorem~\ref{theoVMG}, we have $\lambda:=(1,0)'\in\DDD_Y$ for $\mu=(0,0)'$, $\Sigma:=\diag(1,1)$ and
\[T\sim \alpha\Gamma_S^2(1,2,(1,1)')=\MMM\Gamma^2_S\Big((1,1,1)',
\left(\begin{array}{ccc}
1/2&0&1/2\\
0&1/2&1/2
\end{array}
\right)
\Big)\,,
\]
but also,  recalling~\eqref{defMlambda},
\[M_{(1,0)'}=(M^{(1,0)'}_1,M^{(1,0)'}_2,M^{(1,0)'}_3)=
\left(\begin{array}{ccc}
2/3&0&2/3\\
0&1/2&2/3
\end{array}
\right)
\,.
\]
Under the Escher transform, the first component of $T$ reads $T_1\eqd(2/3) G_1+(2/3)G_3$ for independent $G_1,G_3\sim\Gamma_S(1)$ by~\eqref{defmultigamma}. However,
as implied by Lemma~\ref{lemgamma}, $(2/3) G_1+(2/3)G_3\sim\Gamma_S(2,3/2)$, and $T_1$ cannot be a standard Gamma process. The associated $V\alpha\Gamma^d_S$-class of subordinated Brownian motions is thus not closed under the Esscher transformation in the interpretation of Theorem~\ref{theoEsscherVGGC}.\\[1mm]
\noindent{\bf Guillaume's Subordinator.} Guillaume~\cite{Gu13} extends Semeraro's $\alpha\Gamma^d$-class as follows:
let $\alpha_*=(\alpha_1,\dots,\alpha_d)',
a_*=(a_1,\dots,a_d)',\beta_*=(\beta_1,\dots,\beta_d)'\in(0,\infty)^d$, $c_1,c_2>0$. Let
$S_1,\dots, S_{d+1}$ be independent
such that
\[
S_k\,\sim\,\Gamma_S(a_k,\beta_k)\,,\;\;1\le k\le d\,,\;\quad\quad
S_{d+1}\,\sim\,\Gamma_S(c_1,c_2)\,.\]
We refer to $T$ as a {\em ${\cal G}\Gamma^d$-subordinator}, in brief $T\sim {\GGG}\Gamma^d_S(\alpha_*,a_*,\beta_*,c_1,c_2)$, provided
$T\eqd (T_1,\dots,T_d)'$ with $T_k:= S_k+\alpha_k S_{d+1}$.

With $S_1,\dots,S_{d+1}$ as above, introduce independent standard Gamma subordinators
$G_1\sim \Gamma_S(a_1) ,\dots,G_d\sim\Gamma_S(a_d),G_{d+1}\sim \Gamma_S(c_1)$ by setting
\[G_k:=\frac {\beta_k}{a_k} S_k\,,\quad 1\le k\le d\,,\quad\quad G_{d+1}:=\frac {c_2}{c_1} S_{d+1}\,.\]
We conclude that  $T\sim \MMM\Gamma^d_S(b_*,M(\alpha_*,a_*,\beta_*,c_1,c_2))$, where in our notation, $b_*=(a_1,\dots,a_d,c_1)'\in(0,\infty)^{d+1}$ and \begin{equation}\label{GuillaumeM}M(\alpha_*,a_*,\beta_*,c_1,c_2)\,:=\,
\Big( \diag\left(a_1/\beta_1,\dots,a_d/\beta_d \right)\,,\;
(c_1/c_2)\alpha_*\Big)\quad\in[0,\infty)^{d\times (d\!+\!1)}_*\,.\end{equation}
Further, observe that
\begin{eqnarray*}
\lefteqn{\{{\GGG}\Gamma^d_S(\alpha_*,a_*,\beta_*,c_1,c_2):\;\alpha_*,a_*,\beta_*\in(0,\infty)^d, c_1,c_2>0\}}&&\\
&=&\{\MMM\Gamma^{d}_S(b_*,\diag(x_*),y_*)):\;x_*,y_*\in(0,\infty)^d,b_*\in(0,\infty)^{d+1}\}\,.
\end{eqnarray*}
By Part~(c) of Theorem~\ref{theoVMG}, the $V{\cal G}\Gamma^d$-class of subordinated Brownian motions
is, thus, closed under the Esscher transformation in Theorem~\ref{theoEsscherVGGC}.

Unlike the $\alpha\Gamma^d$-subordinator, a given $\GGG\Gamma^d$-subordinator does not need to have Gamma marginals, and we clarified this in Part~(a) of Lemma~\ref{lemsubgammaIII}. By Part~(b) of Lemma~\ref{lemsubgammaIII}, contemplating~\eqref{GuillaumeM}, a $\GGG\Gamma^d$-subordinator and, thus, any $\alpha\Gamma^d$-subordinator is also a $\Gamma^d_S$-subordinator, concluding settlement of our diagram in  Figure~\ref{diagramsubord}, also recalling the chain of inclusions
$\alpha\Gamma^d\subseteq\GGG\Gamma^d_S\subseteq\MMM\Gamma_S^d$.
\section{Applications}\label{secapplic}
We are primarily concerned with demonstrating how our $V\MMM\Gamma^d$-subclass can be applied, in particular, to price multi-asset options.
The $V\MMM\Gamma^d$-subclass, as we showed, contains other popular models, such as the multivariate VG~\cite{MaSe90}, the Semeraro $\alpha VG$~\cite{Se08}, and Guillaume's extension~\cite{Gu13}.

In Subsection~\ref{subsecmarketmodel} a market model using the $V\MMM\Gamma^d$-process
is introduced, and we give explicit formulae for the expected value of the
 $k$-dimensional log-price process and its covariance matrix, and for the
 expected value of the price process itself. This allows us to tabulate values of these quantities for a specific parameter set which we will use to illustrate the results. The corresponding densities are calculated using the formula for the characteristic function given in \eqref{charGVGmulti} of Theorem~\ref{theoGVGmulti} and displayed in Figure~\ref{fig1}. The parameters required to make the Esscher transform an equivalent martingale measure linking the real world and risk neutral dynamics are derived in Proposition \ref{EsscherProp2} of Subsection~\ref{subsecriskneutral}.
As an example, pricing of four kinds of two-asset options, specifically, European and American best-of and worst-of put options, can then be operationalised as we demonstrate in Subsection~\ref{subsecoption}. The exact form of the L\'evy measure as given in Theorem~\ref{theoVMG}~(b) is an essential ingredient here.
\subsection{A $V\MMM\Gamma^d$-Market Model}\label{subsecmarketmodel}
We employ the $V\MMM\Gamma^d$-process to model the log-prices of risky assets of a financial market.
Potentially latent risk factors are described by a process $Y\sim V\MMM\Gamma^d(b_*,M,\mu,\Sigma)$-process with respect to a given stochastic basis $(\Omega,\FFF,\{\FFF_t\},P)$.
The risk factors drive a $k$-dimensional price process $S$ with $S_i(t) = S_i(0) \, e^{R_i(t)}$, for $t \ge 0$ and $i = 1,...,k$, with $k$-dimensional log-price process $R=(m-q+\kappa)I+A Y=\{R(t):t\ge 0\}$ given by
\begin{equation}\label{defRriskprocess}
    R(t)\,=\,\left(m - q + \kappa\right)\, t + A\, Y(t)\,=\, (m - q + \kappa) \, t +  X(t),\; t\ge 0\,,
\end{equation}
where $m \in \RR^k$ is the expected total return rate of the assets, $q\in\RR^k$ is the dividend yield of the assets,
$A \in \mathbb{R}^{k\times d}$ with rows $A^i \in \RR^d$ satisfying ${A^i}^\prime \in {\cal D}_Y$, $i =1,...,k$, determines the factor loading of the corresponding log-return process, and $\kappa\in \RR^k$ is an adjustment vector given by $\kappa_i = -\log E  e^{X_i(1)} = -\log E e^{\langle {A^i}^\prime, Y(1)\rangle}$ such that $E S_i(t) = S_i(0) \, e^{(m_i -q_i)\,t}$, $t \ge 0$, $i=1,...,k$.
Recall $I:[0,\infty)\to[0,\infty)$ denotes the identity function.
Proposition~\ref{VMGMomentsProp} gives formulae for the moments of $R(t)$ 
and the explicit form of the adjustment vector $\kappa$.
\begin{remark}\label{remA}
The dependence structure of the risk factor process $Y$ is limited, as $\Sigma$ has to be a diagonal matrix in order that we remain in the class of L\'evy processes.
The matrix $A$ maps those risk factors to specific asset prices and generates a richer and perhaps more realistic dependence structure, for similar arguments and setup see~\cite{LS10,Ma12b,Se08}.
Accordingly, $A Y$ and $R$ are {\em not} necessarily $V\MMM\Gamma^k$-processes, but are of course L\'evy processes.\halmos
\end{remark}

\begin{proposition}\label{VMGMomentsProp}
Let $t\ge 0$ and $R$ as in~\eqref{defRriskprocess} with $n=\dim b_*$. Then:
\\
(a) $E R(t) = \left(m - q + \kappa + A \sum_{l=1}^n \mu \diamond M_l \right) t, \quad t \ge 0$.
\\
(b) {\em Cov}$(R(t)) = A \, \left[\sum_{l=1}^n \left( \frac{1}{b_l}(\mu\diamond M_l) (\mu\diamond M_l)^\prime + \Sigma \diamond M_l\right)\right]{A}^\prime\, t,  \quad t \ge 0$.
\\
(c)~$\kappa_i = \sum_{l=1}^n b_l \, \log\left\{ \left({b_l - \langle  \mu \diamond M_l, A^i{}^\prime \rangle - \frac{1}{2}\| A^i{}^\prime\|^2_{\Sigma\diamond M_l} } \right)/ b_l \right\}\,, \quad i = 1,...,k$.
\end{proposition}
\noindent{\em Proof.} This follows by differentiating the Laplace transform (see~\eqref{LaplaceVMG} in  Theorem~\ref{theoVMG}).
\halmos
\begin{table}[hb!]
\centering
\begin{tabular}{c|ccccc}
\hline \hline
$\rho$ & $E R_1(1)$ & $E R_2(1)$ & $\mbox{Var}( R_1(1))^{\frac{1}{2}}$ &  $\mbox{Var}( R_2(1))^{\frac{1}{2}}$ & $\mbox{Cor}( R_1(1), R_2(1))$ \\
\hline\hline
\phantom{-}0.30\phantom{-}  & 0.0917 & 0.0782 & 0.1296 & 0.2104 & 0.3651 \\
\phantom{-}0.00\phantom{-}  & 0.0921 & 0.0780 & 0.1260 & 0.2114 & 0.0329 \\
          -0.30\phantom{-}  & 0.0919 & 0.0785 & 0.1276 & 0.2092 &-0.3076
\\\hline \hline
\end{tabular}
\caption{\it Expected value, volatility and correlation of $R(1)$ for $A = (1, \rho; \rho, 1)^{0.5}$, $\rho \in \{-0.30,0,0.30\}$, $Y\sim V\MMM\Gamma^d(b_*,M,\mu,\Sigma)$ with parameters $n = 3$, $d = k = 2$, $m = (0.1, 0.1)$, $q = (0,0)$, $b_* = ( 5, 5, 10)'$, $M = ( 0.5, 0, 0.5; 0, 0.5, 0.5)$, $\mu = ( -0.14, -0.25)$, $\Sigma = {\rm diag}(0.0144, 0.04 )$.}\label{moments}
\end{table}
We investigate the distribution of $R$ for parameters: $d=k=2$,  $m=(0.1, 0.1)$, $q=(0,0)$,
$b_*=( 5, 5, 10)'$, $M=( 0.5, 0, 0.5; 0, 0.5, 0.5)$, $\mu=( -0.14$,\\$ -0.25)$, $\Sigma={\rm diag}( 0.0144, 0.04 )$ and $A = (1, \rho; \rho, 1)^{0.5}$ with $\rho \in \{-0.3, 0, 0.3\}$.
Table~\ref{moments} states the expected value, volatility (square root of variance), and correlation of $R(1)$, for $\rho \in \{-0.3, 0, 0.3\}$.
These numbers facilitate a better understanding of the potentially abstract model parameters and serve as a basis for comparison when the Esscher transform is discussed in Subsection~\ref{subsecriskneutral}.
The expected values for both coordinates are below $m=(0.1,0.1)$ and are robust when varying $\rho$.
The expected value of the first coordinate becomes maximal for $\rho = 0$ whereas for the second coordinate the relationship is inverted.
This effect is determined by the term $A \sum_{l=1}^n\mu \diamond M_l$ in Proposition~\ref{VMGMomentsProp}~(a).
A similar behavior can be observed for the volatilities, however, here the roles of the coordinates are exchanged.
Most notably, the correlation differs considerably from the dependence parameter $\rho$.
The main driver of this difference is the first component $A \, \left[\sum_{l=1}^n \frac{1}{b_l}(\mu\diamond M_l) (\mu\diamond M_l)^\prime \right]{A}^\prime$ in Proposition~\ref{VMGMomentsProp}~(b).
Depending on the sign of the entries of $A\mu$ this term increases or decreases the correlation.
For $\rho\in\{-0.30,0,0.30\}$, $A\mu$ has negative entries in both coordinates, consequently increasing the correlation above $\rho$.
This effect weakens when decreasing the dependence parameter $\rho$.

\begin{figure}[hpbt!]
\begin{center}
\center{${ }$ \phantom{--} $\rho = 0.30$ \phantom{-} \qquad\qquad $\rho = 0.00$ \qquad\qquad $\rho = -0.30$}

\vspace{1.00cm}

\flushleft{$t=0.01$}

\vspace{2.50cm}

\flushleft{$t=0.25$}

\vspace{-5.50cm}

\vspace{-2.5cm}

\hspace{-0.0cm}
\includegraphics[width=0.45\textwidth]{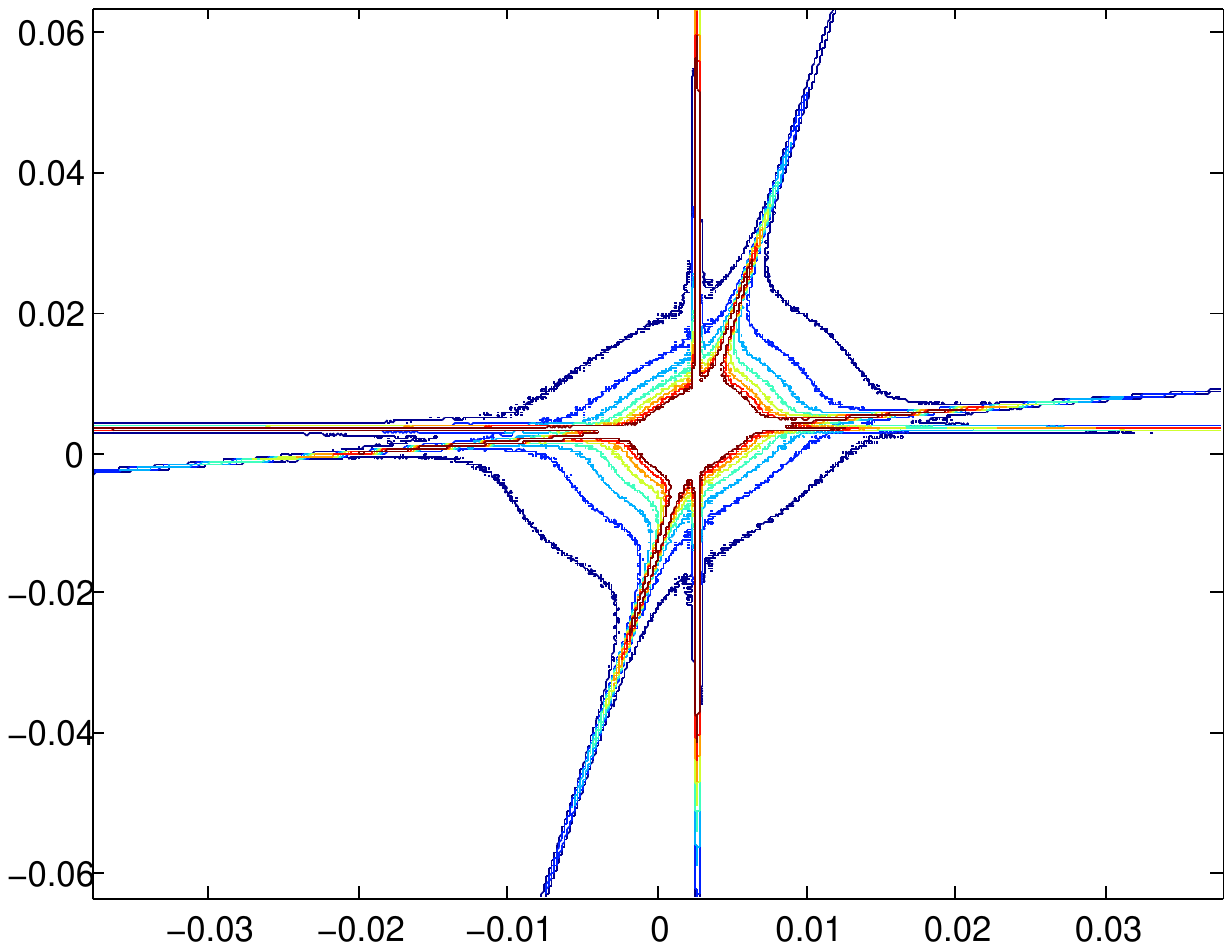}
\hspace{-2.75cm}
\includegraphics[width=0.45\textwidth]{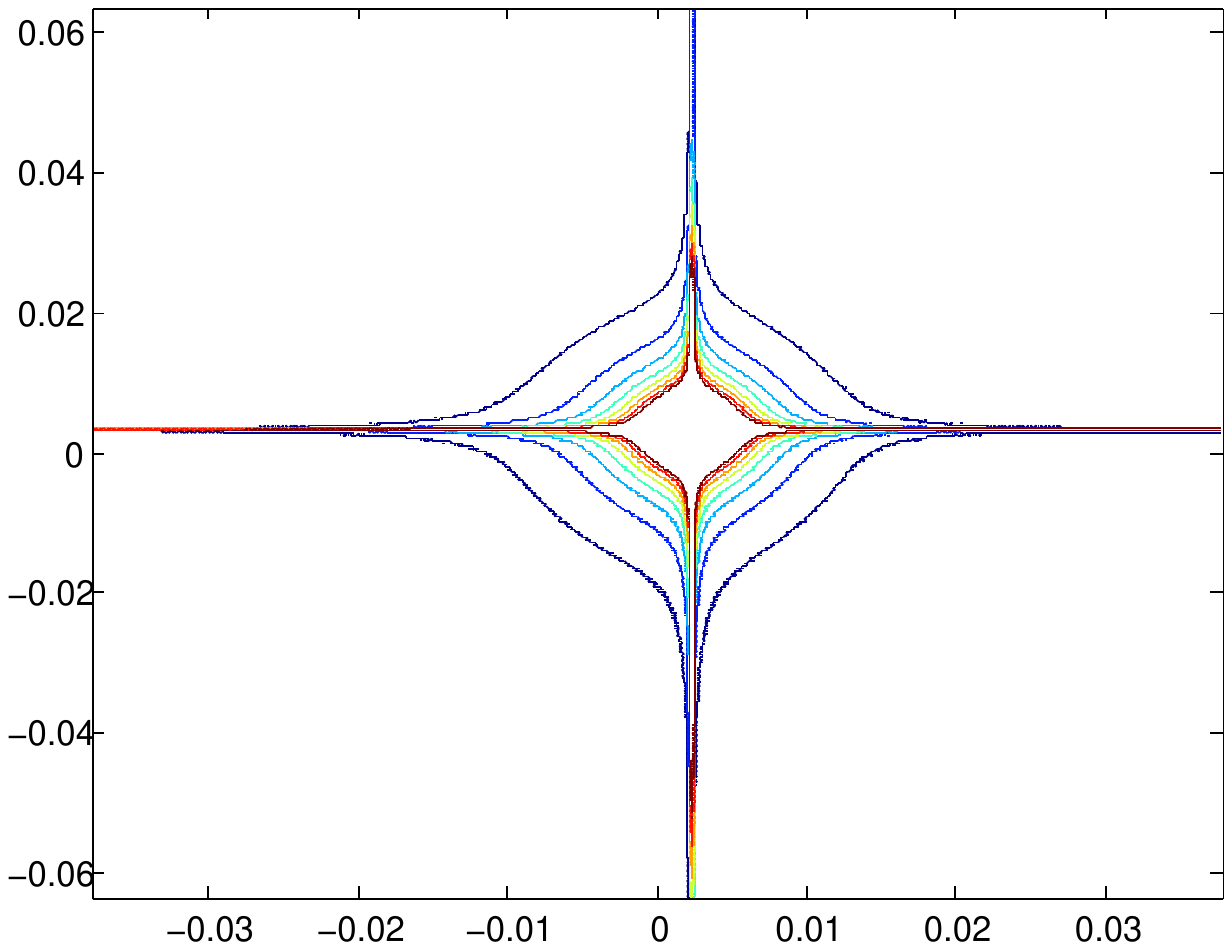}
\hspace{-2.75cm}
\includegraphics[width=0.45\textwidth]{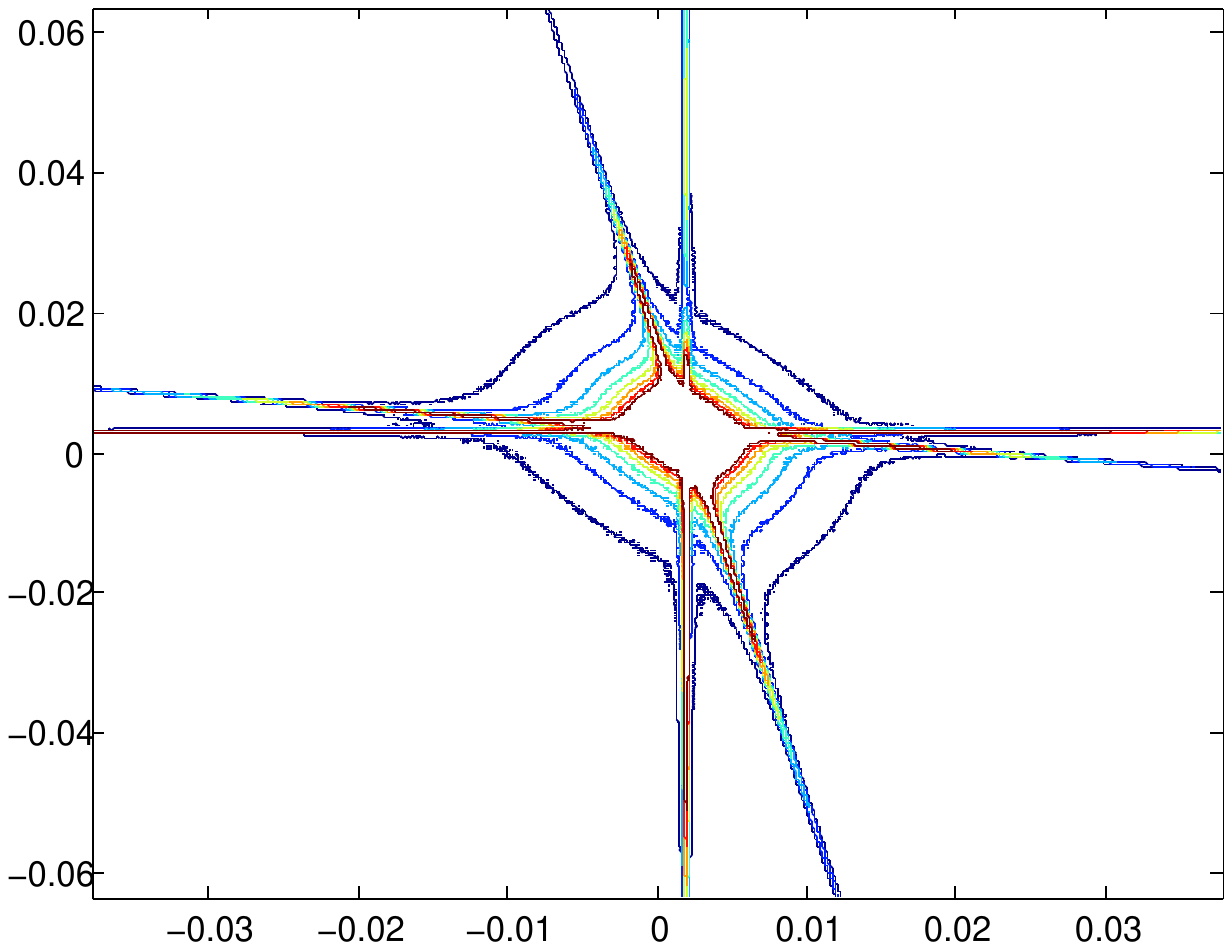}

\vspace{-5.5cm}

\hspace{-0.0cm}
\includegraphics[width=0.45\textwidth]{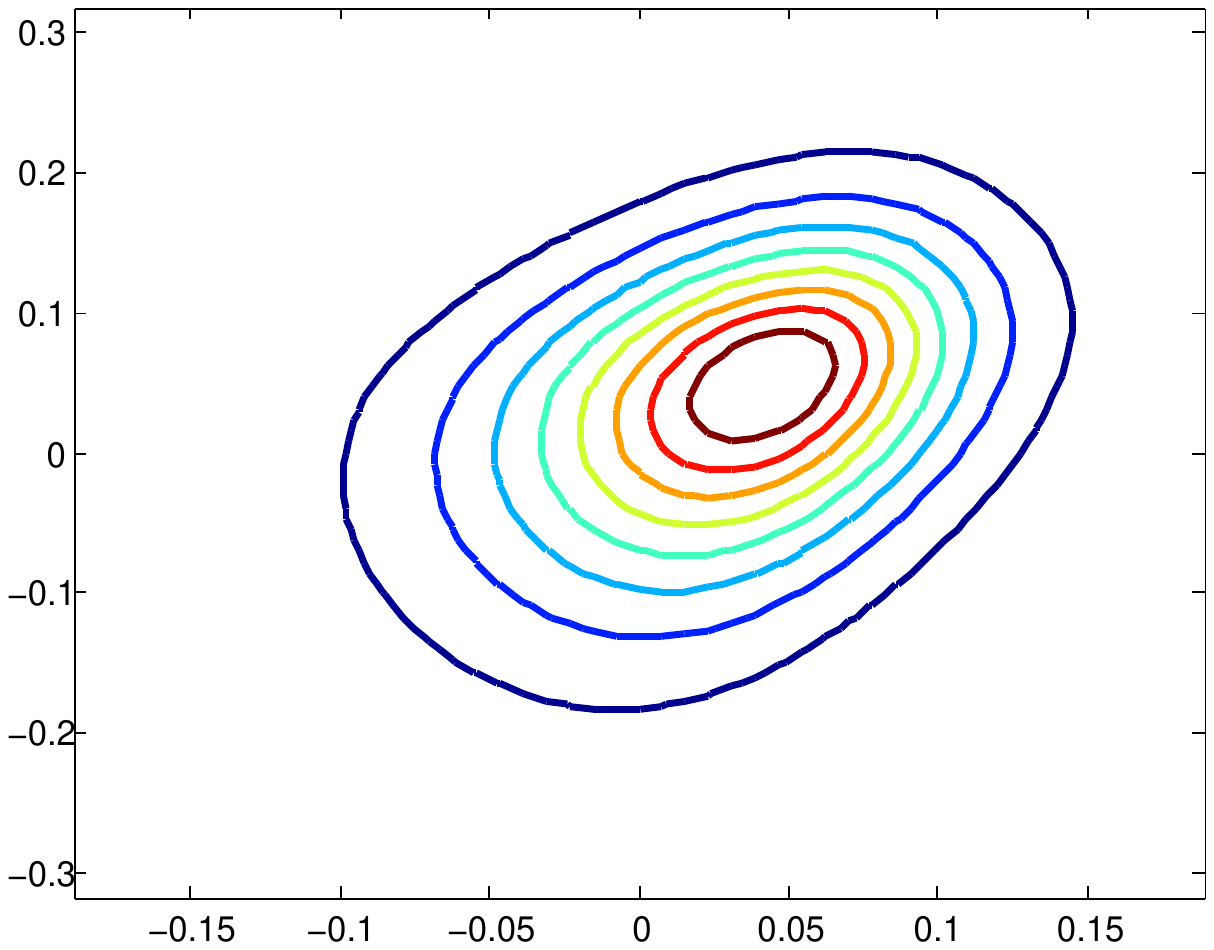}
\hspace{-2.75cm}
\includegraphics[width=0.45\textwidth]{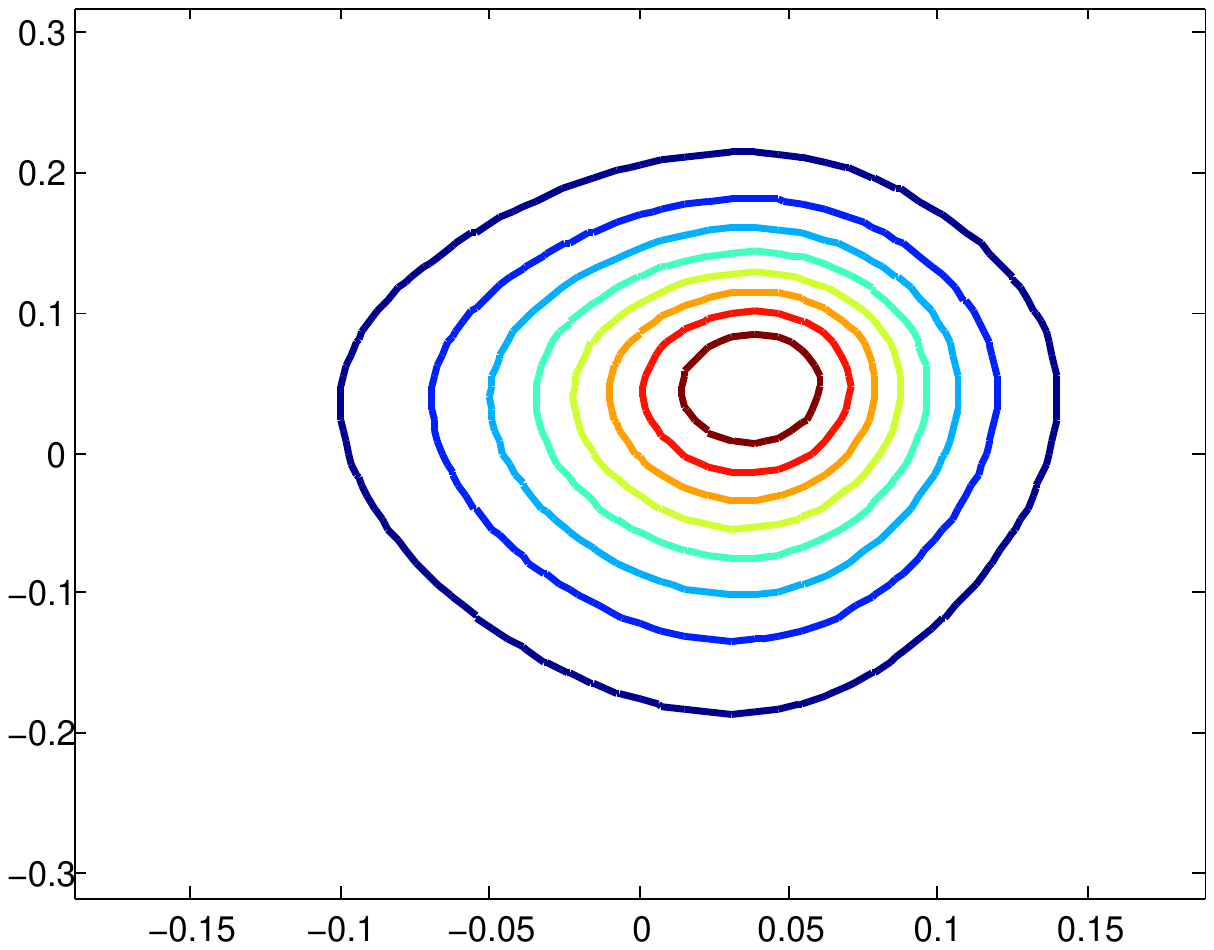}
\hspace{-2.75cm}
\includegraphics[width=0.45\textwidth]{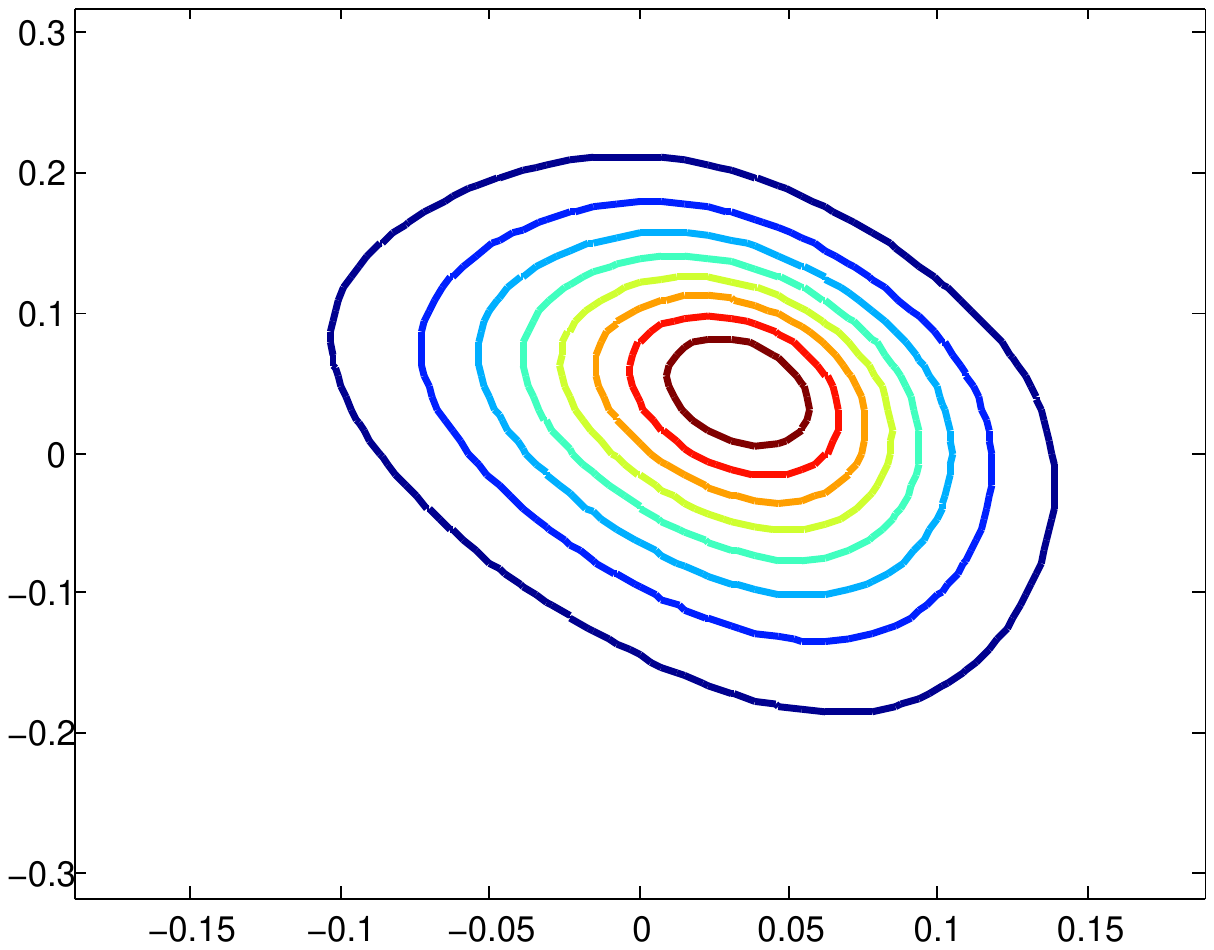}

\vspace{-3cm}

\caption{\it Density level lines of $R(t)= (m-q+\kappa)t+AY(t)$ for $t\in\{0.01,0.25\}$, $A = (1, \rho; \rho, 1)^{0.5}$, $\rho \in \{-0.30,0,0.30\}$, $Y\sim V\MMM\Gamma^d(b_*,M,\mu,\Sigma)$ with parameters the same as for Table~\ref{moments}.}\label{fig1}
\end{center}
\end{figure}

Figure~\ref{fig1} illustrates the density of $R$ for $t\in\{0.01,0.25\}$ when varying $\rho\in\{-0.30,0,0.30\}$.
The densities are obtained numerically from the characteristic function given in \eqref{charGVGmulti} of Theorem~\ref{theoGVGmulti} using fast Fourier inversion. For $t=0.01$, the superposed processes $A \mu \diamond T$ dominate $A \, \Sigma^{1/2} \hat B \circ_{d} T$,
where $T\sim \MMM\Gamma_{S}^d(b_*,M)$ and $\hat B$ is $d$-dimensional standard Brownian motion.
For $\rho = 0$, most of the probability mass is located near the $x$- and $y$-axes.
For $\rho = 0.30$, additionally mass appears around two straight lines in the first and third quadrants (positive dependence).
For $\rho = -0.30$, additionally mass appears around two straight lines in the second and fourth quadrants (negative dependence).
For $t=0.25$, the density is close to normal with nearly elliptical level lines.
Note, though, that for $\rho=0$ the density is not symmetric but skewed towards the left and lower values.

\begin{remark}\label{remmargins} A desirable property of a parametrisation of a multivariate distribution is to be able to distinguish between parameters describing marginal distributions, and parameters describing the dependence.
For the $V\MMM\Gamma^d$, however, this is in general not possible.
Each parameter appears in at least one marginal distribution.
This is a consequence of the fact that the family of Gamma distributions is not stable under convolution, except for singular cases;
see Lemma~\ref{lemgamma}. These are the cases analysed by~\cite{Se08}.
See also \cite{Ka09} for correlating L\'evy process and related applications.
\halmos
\end{remark}

%
%

\subsection{Risk-Neutral Valuation via Esscher Transform}\label{subsecriskneutral}
Option pricing requires a risk-neutral measure as the basis for risk-neutral valuation.
In the general L\'evy process setting, such a  measure is not guaranteed to exist and further, if it exists it is in general not unique.
But in Part~(c) of Theorem~\ref{theoVMG} we showed that the $V\MMM\Gamma^d$-class is invariant under an Esscher transformation, and here
we follow common practice by adopting the Esscher transformation for identifying a risk-neutral measure, see~\cite{CT,GS94,Ta11}.

For the processes $R,X,Y$ in~\eqref{defRriskprocess} and
$h\in\DDD_R=\DDD_X=\DDD_{AY}$ the Esscher transform is given by (see~\eqref{defEsscher})
\begin{align}\label{EsscherDeft}
    \frac{\rmd Q^R_{h,t}}{\rmd P} = \frac{e^{\langle h, R(t)\rangle}}{{E}_P[e^{\langle h, R(t)\rangle}]}
    = \frac{e^{\langle h, X(t)\rangle}}{{E}_P[e^{\langle h, X(t)\rangle}]}
    = \frac{e^{\langle A'h, Y(t)\rangle}}{{E}_P[e^{\langle {A'h}, Y(t)\rangle}]}\, ,
    \quad t\ge 0\,,
\end{align}
such that, with $h\in\DDD_R=\DDD_X=\DDD_{AY}$,
\begin{align}\label{Esscherids}
    \frac{\rmd Q^R_{h,t}}{\rmd P} = \frac{\rmd Q^X_{h,t}}{\rmd P}=\frac{\rmd Q^Y_{A'h,t}}{\rmd P}
      \, ,
    \quad \text{for } t\ge 0\,.
\end{align}
By Part~(c) of Theorem~\ref{theoVMG}, as $\DDD_R=\DDD_X=\DDD_{AY}$, we observe that
\begin{align*}
    \DDD_R=& \Big\{ h \in\mathbb{R}^k:  \langle\mu \diamond M_l,A'
     h\rangle + \frac{1}{2}\| A' h\|^2_{\Sigma\diamond M_l} < b_l,\quad 1\le l \le n\Big\}\,.
\end{align*}
Also, by replacing $\lambda$ with $A'h$ in Theorem~\ref{theoVMG}, it follows from~\eqref{Esscherids} that
\[\{Y(s):0\le s\le t\}|Q^R_{h,t}\quad\sim\quad V\MMM\Gamma^d(b^h_*,M^h,\mu^h,\Sigma^h)\,,\quad h\in\DDD_R\,,\;t\ge 0\,,\]
with $b^h_*  =  b_*$, $\mu^h =  \mu + \Sigma \, A' h$, $\Sigma^h  = \Sigma$, and
\begin{align*}
    M^h_l & = \frac{b_l}{b_l - \langle\mu \diamond M_l,A' h\rangle - \frac{1}{2}\| A' h\|^2_{\Sigma\diamond M_l}} \, M_l\, , \qquad 1 \le l\le n\,.
\end{align*}
Next, we summarise risk-neutral pricing under the Esscher transform:
\begin{proposition}\label{EsscherProp2}Assume $h^\star\in\RR^k$ such that $h^\star, \eeee_i\!+\!h^\star \in \DDD_R=\DDD_X=\DDD_{AY}$, $1\!\le\! i\! \le\! k$.
Then, for the market with price process $S_0 = e^{rI}$ and $S_i = S_i(0)\, e^{R_i}$ with $S_i(0) \in\RR^+$, $1\le i \le k$,
the Esscher transform $Q^R_{h^\star}$ is an equivalent martingale measure with respect to the numeraire $S_0$:~$Q^R_{h^\star,\,T}\sim P$ and $e^{q_iI} S_i/S_0$ are $Q^R_{h^\star,\,T}$-martingales, for $1\le i\le k$ and $T>0$ if and only if
\begin{align}\label{EsscherRN}
    m_i -r & =  \Lambda_{AY(1)}(\eeee_i) + \Lambda_{AY(1)}(h^\star) -\Lambda_{AY(1)}(\eeee_i+h^\star)  \, , \quad \text{for } 1\le i \le k\,,
\end{align}
where $\Lambda_X$ is the cumulant-generating function of an $\mathbb{R}^d$-valued random variable $X$, i.e. $\Lambda_X(u) = \log E e^{\langle u, X\rangle}$, $u \in  \{v \in \mathbb{R}^d: E e^{\langle v, X\rangle} < \infty \}$.
\end{proposition}
\noindent{\it Proof.}~Let $h \in \DDD_{AY}$ such that $h + \eeee_i \in\DDD_{AY}$, for $1\! \le\! i\! \le k$.
Then $Q_h:=Q^R_{h,T}$ is well-defined and $E_{Q_h} |e^{q_i t} S_i(t)/S_0(t)|\! <\! \infty$, for $1\!\le\! i\! \le\! k$ and $0\!\le\! t\! \le\! T$.
Note that $e^{q_iI} S_i/S_0$ is the exponential of a L\'evy process, under both $P$ and $Q_{h,t}^R$, and thus for $1\le i \le k$ and $0 \le t \le T$ it is the case that
\begin{align*}
    E_{Q_{h}}\left[ \left. e^{q_i T} S_i(T)/S_0(T) \right| {\cal F}_t\right]
    & = \frac{e^{q_i t } S_i(t)}{S_0(t)} \, \left( e^{q_i} E_{Q_{h}}\left[ \frac{S_i(1)/S_i(0)}{S_0(1)/S_0(0)} \right]\right)^{T-t}
    \\
    & = \frac{e^{q_i t } S_i(t)}{S_0(t)} \, \left( e^{m_i +\kappa_i-r} \,\frac{E_{P}\left[ e^{\langle \eeee_i, AY(1)\rangle} \,e^{\langle h, 
    {X(1)}
    \rangle}\right]}{E_{P}\left[ e^{\langle h,
     X(1)\rangle} \right]}\right)^{T-t}
    \\
    & = \frac{e^{q_i t } S_i(t)}{S_0(t)} \, \left( e^{m_i +\kappa_i-r} \,\frac{E_{P}\left[ e^{\langle \eeee_i +h, AY(1)\rangle} \right]}{E_{P}\left[ e^{\langle h, AY(1)\rangle} \right]}\right)^{T-t}
    \\
    & = \frac{e^{q_i t } S_i(t)}{S_0(t)} \, e^{(m_i +\kappa_i-r + \Lambda_{AY}(\eeee_i+h) - \Lambda_{AY}(h))(T-t)}\,.
\end{align*}
Recall $\kappa_i = -\log E e^{\langle {A^i}^\prime, Y(1)\rangle} = -\log E e^{\langle \eeee_i , AY(1)\rangle} = -\Lambda_{AY}(\eeee_i)$ to see that $e^{q_i I} S_i/S_0$ is a $Q_h$-martingale,
$i = 1,...,k$,
if and only if $h$ satisfies \eqref{EsscherRN}.\halmos
\begin{remark} The parameter $h^\star$ is called the {\em Esscher parameter.} For general exponential L\'evy market models,
Theorems 4.4--4.5 in \cite{KS02} ( also see~\cite{RW15}, their Theorem~2.6) state that $h^\star$ is unique, provided the driving L\'evy process does not degenerate under $P$ in the sense of Definition~24.16 of~\cite{s}.~An application of this result yields that our market model~\eqref{defRriskprocess} admits a unique $h^\star$, provided $\rank(A)\ge k$, $\rank(M)\ge d$ and $\det\Sigma>0$.\halmos
\end{remark}

\begin{table}[hb!]
\footnotesize
\centering
\begin{tabular}{c|cccc}
\hline \hline
 & $\rho = \phantom{-}0.30$ & $\rho = \phantom{-}0.00$ & $\rho = - 0.30$ \\
\hline\hline
$h^\star$        & $(-2.5626, -0.5351)^\prime$ & $(-2.9662, -1.0410)^\prime$ & $(-3.8416, -1.8390)^\prime$ \\ \hline
$\mu^{h^\star}$  & $(-0.1776, -0.2867)^\prime$ & $(-0.1827, -0.2916)^\prime$ & $(-0.1907, -0.2994)^\prime$ \\
$M^{h^\star}$    & $\left(\begin{array}{cc} 0.5217 & 0 \\ 0 & 0.5126 \\ 0.5171 & 0.5171\end{array}\right)^\prime$
                 & $\left(\begin{array}{cc} 0.5251 & 0 \\ 0 & 0.5145 \\ 0.5198 & 0.5198\end{array}\right)^\prime$
                 & $\left(\begin{array}{cc} 0.5309 & 0 \\ 0 & 0.5176 \\ 0.5241 & 0.5241\end{array}\right)^\prime$ \\ \hline
$E_{{h^\star}}R(1)$  & $(0.0408, 0.0268)'$ & $(0.0412, 0.0264)'$ & $(0.0409, 0.0266)'$ \\
$\mbox{Var}^{1/2}_{{h^\star}}R_1(1)$  & $0.1365$ & $0.1334$ & $0.1359$ \\
$\mbox{Var}^{1/2}_{{h^\star}}R_2(1)$  & $0.2178$ & $0.2195$ & $0.2185$ \\
$\mbox{Cor}_{{h^\star}}(R_1,R_2)$     & $0.3751$ & $0.0492$ & $-0.2864\phantom{-}$
\\ \hline \hline
\end{tabular}
\caption{\it Esscher parameter and resulting basic statistics for $A = (1, \rho; \rho, 1)^{0.5}$, $\rho \in \{-0.30,0,0.30\}$, $r=0.05$, $Y\sim V\MMM\Gamma^d(b_*,M,\mu,\Sigma)$ with parameters the same as for Table~\ref{moments}.}\label{tabEsscher}
\end{table}

Next we set the interest rate to $r = 0.05$ and keep the remaining model parameters as in Subsection~\ref{subsecmarketmodel}.
The resulting Esscher parameter, the adjusted risk-neutral parameters and some basic statistics are provided in Table~\ref{tabEsscher}.
The first row indicates the three different scenarios, i.e. $\rho \in \{-0.30,0,0.30\}$.
In the second row the Esscher parameter $h^\star$ is seen to have negative components that are increasing in $\rho$.
The sign of the components of $h^\star$ is as expected, since the model under $P$ corresponds to a ``bullish'' market with expected return rates of $m_1 = m_2 = 0.1$ exceeding the risk free rate $r=0.05$, and $h^\star$ has to counterbalance this effect.
The third row gives the transformed parameter $\mu^{h^\star}$ which tends to be lower than the original parameter under $P$ and is increasing in $\rho$ as well.
The matrix distributing the Gamma subordinators to the coordinates $M^{h^\star}$ is displayed in the fourth row.
The elements are all greater than those of $M$ and the more negative the dependence parameter $\rho$ becomes the stronger is this effect.
The resulting characteristics of the distribution are displayed in Rows~5 to~8.
These numbers can be compared to the numbers under $P$ in Table~\ref{moments}.
The expected values of $R(1)$ under the Esscher martingale measure are lower than under~$P$.
The volatilities increase across the board by nearly $1\%$.
For the correlation the same can be observed; an increase of about $1\%$ is found when comparing the Esscher numbers to the original numbers under~$P$.
Summarising, volatilities and correlations increase when we change from $P$ to $Q^{h^\star}$.
Thus under the pricing measure $Q^{h^\star}$ risk in the form of volatilities requires a higher risk premium than would be anticipated under $P$, e.g., when pricing a call or put option.
Further, diversification effects are less pronounced under the pricing measure, e.g., requiring a higher premium for basket options.

\subsection{Pricing Best-of and Worst of Put-Options}\label{subsecoption}
The financial market model presented above can capture a wide range of dependencies between different asset prices.
As an illustration we price some cross-dependence sensitive options of both European and American styles.
European options can be conveniently priced by Fourier methods~\cite{CM99}.
Thus, we can draw on the results provided in Theorem~\ref{theoGVGmulti} to compute European option prices.
Pricing American options can be carried out by finite difference methods, discretising the respective pricing partial integro-differential equations, or by using tree-based methods. See~\cite{HS09} for a recent survey on numerical methods in exponential L\'evy process models. Both methods require formulae for the L\'evy measure that we provided in Theorem~\ref{theoVMG}.

As an example we consider best/worst-of put options with respective early exercise values
\begin{align}\label{bop}
    \chi_{{\rm bop}, k}(t) & = \left(K - \bigvee_{i=1}^k S_i(t)\right)^+\,,\quad
    \chi_{{\rm wop}, k}(t) = \left(K - \bigwedge_{i=1}^k S_i(t)\right)^+\, ,
\end{align}
for $0 \le t \le T$, where $T$ is the maturity date and $K \in\mathbb{R}^+$ the exercise price.

The risk-neutral parameters are: $n = 3$, $d = k = 2$, $b_* = ( 5, 5, 10)'$, $M = ( 0.5, 0, 0.5; 0, 0.5, 0.5)$, $\Sigma = {\rm diag}( 0.0144, 0.04 )$, $\mu = ( -0.14, -0.25)$, $m = (0.1, 0.1)$, $q = (0,0)$ and $A = (1, \rho; \rho, 1)^{0.5}$ with $\rho \in \{-0.3, 0, 0.3\}$.
Note that we have set here $r=0.1$ in contrast to Subsection~\ref{subsecriskneutral}, resulting in $h^\star = 0$ and $Q^{h^\star} = P$.
This allows us to interpret the option price dependencies on the parameter $\rho$ without confounding this with effects of the Esscher transform on the option premium.
To compute American option prices we use the tree approach as outlined in~\cite{Kae09,KMS13}, based on~\cite{MSZ06}.
The European option prices are obtained as a byproduct of this procedure.

\begin{table}[hpbt!]
\centering
\begin{tabular}{cc|cc|cc}
\hline \hline
\multirow{2}{*}{$\rho$} & \multirow{2}{*}{$K$} & \multicolumn{2}{c}{Best-of put price} & \multicolumn{2}{|c}{Worst-of put price} \\
&& European & American & European & American\\
\hline\hline
0.3 & 90 & 0.04 
                &  0.05 
                        & 0.75 & 0.81\\
0.3 & 95 & 0.18 & 0.24 & 1.76 & 1.90\\
0.3 & 100 & 0.71 & 1.06 & 3.74 & 4.03\\
0.3 & 105 & 2.17 & 5.00 & 7.00 & 7.49\\
0.3 & 110 & 4.98 & 10.00\phantom{1} & 11.32\phantom{1} & 11.98\phantom{1}\\\hline
0 & 90 & 0.01 
                & 0.02 
                        & 0.76 & 0.82\\
0 & 95 & 0.09 
             & 0.13 & 1.83 & 1.98\\
0 & 100 & 0.44 & 0.77 & 3.96 & 4.27\\
0 & 105 & 1.63 & 5.00 & 7.48 & 7.96\\
0 & 110 & 4.27 & 10.00\phantom{1} & 12.01 & 12.62\phantom{1}\\ \hline
-0.3 & 90 & 0.00 
                & 0.01 
                        & 0.77 & 0.83\\
-0.3 & 95 & 0.03 
                & 0.06 
                        & 1.85 & 2.01\\
-0.3 & 100 & 0.24 & 0.53 & 4.14 & 4.45\\
-0.3 & 105 & 1.19 & 5.00 & 7.94 & 8.42\\
-0.3 & 110 & 3.66 & 10.00\phantom{1} & 12.63\phantom{1} & 13.20\phantom{1}\\ \hline
\hline
\end{tabular}
\caption{\it Best-of and worst-of put option prices for $T=0.25$, $K \in \{90, 95,100,105,110\}$, $A = (1, \rho; \rho, 1)^{0.5}$, $\rho \in \{-0.30,0,0.30\}$, $r=0.10$, $Y\sim V\MMM\Gamma^d(b_*,M,\mu,\Sigma)$ with parameters the same as for Table~\ref{moments}.
}\label{prices}
\end{table}

The recombining multinomial tree calculation we use has probability weights derived from the L\'evy measure, as provided in Theorem~\ref{theoVMG}.
The option parameters are set to $T=0.25$ and $K\in\{90,95,100,105,110\}$.
The tree models the bivariate process $Y=(Y_1,Y_2)'$
directly, with an exponential transform to obtain the price process.
At each node of the tree the process branches on a regular rectangular $127 \times 127$ grid.
The minimum step sizes are $4.92\times 10^{-3}$ and $8.37\times 10^{-3}$ for $Y_1$ and $Y_2$ respectively.
Prices are then obtained to an accuracy of three significant digits. The time increment is $1.25\times 10^{-3}$.
Run times are reduced by truncating propagation of the tree in its spatial dimensions after one time increment.
Allowing the tree to grow further does not affect the results.

The results are presented in Table~\ref{prices}.
As expected, put options prices are increasing in the exercise price $K$.
Also, the worst-of put option prices exceed the corresponding best-of put option prices, which is consistent with no-arbitrage.
For out-of-the-money options, the early exercise premium is higher for the worst-of put compared to the best-of put.
The early exercise premium for at-the-money options is approximately similar in both cases.
For in-the-money options, the early exercise premium is higher for the best-of put compared to the worst-of put.
The dependence parameter $\rho$ affects the option prices as expected.
The payoff of the best-of put increases the contingency that both price processes fall jointly, thus the option premium is increasing in $\rho$.
The payoff of the worst-of put increases if at least one price process falls, thus the option premium is decreasing in $\rho$.
\section{Proofs}\label{secproof}
\subsection{Polar Decomposition of Measures}\label{subsecpolarrep}
For $\mu,\nu$ being $\sigma$-finite measures, $\mu\otimes \nu$ denotes the corresponding unique $\sigma$-finite product measure.
The trace field of the $d$-dimensional Borel field $\BBB(\RR^d)$ in $A\in\BBB(\RR^d)$ is denoted by $\BBB_A^d$, and
$\mySS^d=\{x\in\RR^d:\|x\|=1\}$ is the unit sphere for a given norm $\|\cdot\|$ on $\RR^d$.
We say that a Borel measure $\mu$ is {\em locally finite relative to} $B\in\BBB(\RR^d)$, provided
$\mu(C)<\infty$ for all compact subsets $C\subseteq B$. Let $\KKK:\mySS^d\times\BBB^1_{(0,\infty)}\to[0,\infty]$ be a {\em locally finite} Borel transition kernel relative to $(0,\infty)$: simultaneously,
$s\mapsto \KKK(s,B)$ is Borel measurable; $B\mapsto \KKK(s,B)$ is a Borel measure, locally finite relative to $(0,\infty)$. There exists a unique measure $\alpha\otimes \KKK:\BBB(\mySS^d)\otimes\BBB((0,\infty))\to[0,\infty]$, locally finite relative to $\mySS^d\times (0,\infty)$ satisfying $(\alpha\otimes \KKK)(A\times B)=\int_A \KKK(s,B)\,\alpha(\rmd s)$ for $A\in\BBB(\mySS^d)$, $B\in\BBB((0,\infty))$ (for example see Exercise~3.24, Chapter~III of \cite{HJI}). Define $\alpha\otimes_p \KKK:\BBB(\RR_*^d)\to[0,\infty]$ as the image of
$\alpha\otimes \KKK$ under $\mySS^d\times (0,\infty)\ni(s,r)\mapsto rs\in\RR^d_*$. By construction, $\alpha\otimes_p \KKK$
is a locally finite Borel measure relative to $\RR^d_*$ satisfing $\int_{\RR^d_*} f(x)\;(\alpha\otimes_p \KKK)(\rmd x)=\int_{\mySS^d}\int_{(0,\infty)} f(rs)\,\KKK(s,\rmd r)\alpha(\rmd s)$ for nonnegative Borel functions $f$.

Next we provide a polar decomposition on $\RR^d_*$ as a disintegration of $\otimes_p$ for Borel measures satisfying additional integrability conditions. The result is directly applicable to L\'evy  and Thorin measures, as in Lemma~\ref{lempolarrec} we may choose $w(r)=r^2\wedge 1$ and $w(r)=(1+\log^- r)\wedge (1/r)$ in view of~\eqref{Piintegrab} and~\eqref{thorinmeasure}, respectively. We omit the proof. It is possible to adapt the arguments in~\cite{BMS06}, Lemma~2.1, and~\cite{Ro90}, Proposition~4.2, respectively.
\begin{lemma}\label{lempolarrec}Assume $0<\int_{\RR^d_*}\! w(\|x\|)\mu(\rmd x)<\infty$ for a Borel measure $\mu$ on $\RR^d_*$
and a continuous function $w:(0,\infty)\!\to\!(0,\infty)$. Then we have:\\[1mm]
(a)~$\mu$ is locally finite relative to $\RR^d_*$ with $\mu(\RR^d_*)\in(0,\infty]$.\\
(b)~There exists a pair $(\alpha,\beta)$ such that, simultaneously,\\
\hspace*{1em}(i)~$\alpha$ is a finite Borel measure on $\mySS^d$;\\
\hspace*{1em}(ii)~$\KKK:\mySS^d\times \BBB((0,\infty))$ is a Borel kernel, locally finite relative to $(0,\infty)$;\\
\hspace*{1em}(iii)~$0<\int w(r)\,\KKK(s,\rmd r)< \infty$ for all $s\in \mySS^d$;\\
\hspace*{1em}(iv)~$\mu=\alpha\otimes_{p} \KKK$.\\
(c)~If $(\alpha',\KKK')$ is another pair, simultaneously satisfying (i)--(iv),  then there exists a Borel function $c:\mySS^d\!\to\!(0,\infty)$ such that $\alpha(\rmd s)=c(s)\alpha'(\rmd s)$ and
$c(s)\KKK(s,\rmd r)=\KKK'(s,\rmd r)$.
\end{lemma}
\subsection{Subordination and Decomposition}\label{subsecsubord}
Let $L^{d,d}(\gamma_X,\Sigma_X,\Pi_X)\subseteq L^d(\gamma_X,\Sigma_X,\Pi_X)$ be the class of L\'evy processes having {\em independent
components}. Let $L^{d,1}(\gamma_X,\Sigma_X,\Pi_X):=L^{d}(\gamma_X,\Sigma_X,\Pi_X)$, $d\in\NN$.

For a Borel measure $\VVV$ on $\RR^d_*$ and $z\in[0,\infty)^d$, we define a Borel measure $\VVV\diamond z$ on $\RR^d_*$ where $(\VVV\diamond z)(A):=\sum_{l=1}^d z_l \VVV(A\cap\AAA_{d,l})$ for a Borel $A\subseteq\RR^d_*$.
Here $\AAA_{1,1}:=\RR$ and $\AAA_{d,l}:=\{x=(x_1,\dots,x_d)'\in\RR^d: x_m=0\mbox{ for }m\neq l\}$, for $d\ge 2$, $1\le l\le d$.
Recalling~\eqref{defotimes1}, introduce $\diamond_{d}:=\diamond$ and $\circ_1:=\circ$. When $z\in[0,\infty)$, $y\in\RR^d$, $\Sigma\in\RR^{d\times d}$ and $\VVV$ is a Borel measure on $\RR^d_*$, we set
$y\diamond_{1} z:=z y$, $\Sigma\diamond_{1}z:=z\Sigma$ and $\VVV\diamond_{1} z:=z \VVV$. Recall~\eqref{defunisub}--~\eqref{defmultisubinproof}.

We collect some formulae for the associated canonical triplets of $X\circ_{k} T$ (see~Theorem~30.1 in~\cite{s} for the univariate subordination; see~Theorem~3.3 in~\cite{BPS01} for the multivariate subordination).
\begin{lemma} \label{lemunifunimultisub}Let $k\in\{1,d\}$.
Let $X\sim L^{d,k}(\gamma_X,\Sigma_X,\Pi_X)$. Let $T\sim S_k(D_T,\Pi_T)$ be independent of $X$. Then we have:\\[1mm]
(a) $X\circ_{k} T\sim L^d(\gamma_{X\circ_{k} T},\Sigma_{X\circ_{k} T},\Pi_{X\circ_{k} T})$ with
\begin{eqnarray*}
\gamma_{X\circ_{k} T}&=& \gamma_X\diamond_{k} D_T+\int_{[0,\infty)^k_*}\,\int_{0<\|x\|\le 1} x\;P(X(s)\in \rmd x)\;\Pi_T(\rmd s)\,,
\\
\Sigma_{X\circ_{k} T}&=&\Sigma_X\diamond_{k} D_T\,,\\
\Pi_{X\circ_{k} T}(\rmd x)&=& (\Pi_X\diamond_{k} D_T)(\rmd x)
+\int_{[0,\infty)^k_*}P(X(s)\in \rmd x)\;\Pi_T(\rmd s)\,.
\end{eqnarray*}
(b) \, For all $t\ge 0$
\[P\big\{(X\circ_{k} T)(t)\in \rmd x\big\}\,=\,\int_{[0,\infty)^k} P(X(s)\in\rmd x)\;P(T(t)\in\rmd s)\,.\]
(c) \, If, in addition, $D_T=0$ and $\int_{[0,1]^k_*} \|t\|^{1/2}\,\rmd\Pi_T(t)<\infty$ then $X\circ_{k} T\sim FV^d(0,\Pi_{X\circ_{k} T})$.
\end{lemma}
In Part (a) of Lemma~\ref{lemunifunimultisub} the dependence of $T$ enters into the formulae in a linear fashion.
As a result, if a process $X$ is independently subordinated by a superposition of independent subordinators then
it can be written (in distribution) as the sum of independent processes:
\begin{proposition}\label{propdecomp} Let $n\ge 1$, $k\in\{1,d\}$ and $X\sim L^{d,k}(\gamma_X,\Sigma_X,\Pi_X)$.

Let $X,T_1,{\dots},T_n$ be independent with $T_l\sim S_k(D_{T_l},\Pi_{T_l})$ for $1\!\le\! l\!\le\! n$.
Let $T:=\sum_{l=1}^n T_k$ and $Y:=X\circ_k T$. Then we have:\\[1mm]
(a)~$T\sim S_k(D_{T},\Pi_T)$ with $D_{T}=\sum_{l=1}^n D_{T_l}$ and  $\Pi_T=\sum_{l=1}^n\Pi_{T_l}$.\\[2mm]
(b)~$Y\sim L^d(\gamma_{Y},\Sigma_Y,\Pi_{Y})$ with $\gamma_{Y}=\sum_{l=1}^n \gamma_{X\circ_{k} T_l}$,
$\Sigma_{Y}=\sum_{l=1}^n \Sigma_{X\circ_{k} T_l}$ and $\Pi_{Y}=\sum_{l=1}^n\Pi_{X\circ_{k} T_l}$.\\[1mm]
(c)~If $X_1,\dots,X_n$ are independent copies of $X$, also being independent of $T_1,\dots,T_n$, then $Y\eqd\sum_{l=1}^nX_l\circ_k T_l$.\\[2mm]
(d)~If, in addition, both $\sum_{l=1}^n\int_{[0,1]_*^k} \|t\|^{1/2}\,\rmd\Pi_{T_l}(t)<\infty$ and
$\sum_{l=1}^n D_{T_l}=0$, then $Y\sim FV^d(0,\Pi_{Y})$ and $X\circ_{k} T_l\sim FV^d(0,\Pi_{X\circ_k T_l})$ for all $1\le l\le d$.
\end{proposition}
\noindent{\it Proof.} (a)~is well known, but can alternatively be deduced from the Laplace transformation. (b)~follows from Part (a), owing to Part (a) of Lemma~\ref{lemunifunimultisub}. (c)~follows from Part (b). (d)~follows from Part (a) as an implication of Part~(c) of Lemma~\ref{lemunifunimultisub}.
\halmos
\subsection{Proofs for Subsection~\ref{subsecVGGC}}~\label{subsecproofVGGC}
\noindent{\em Proof of Theorem~\ref{theoGVG}.}~(a)~Let $Y\eqd B\circ T\sim \VGGC^{d,1}(a,\mu,\Sigma,\TTT)$ where $T,B$ are independent
with $T\sim GCC^1(a,\TTT)$ and $B\sim BM^d(\mu,\Sigma)$. Observe that~\eqref{LaplaceGGCd}
extends to $\lambda\in\CC$ with $\Re\lambda\ge 0$.
This follows from Schwarz's principle of reflection: the proof of Theorem~24.11 of \cite{s} can be adapted to our situation.
Let $\theta\in\RR^d$ and set $\lambda_\theta:=\frac 12 \|\theta\|_\Sigma^2-\rmi \skal \mu\theta$ such that $Ee^{\rmi\skal\theta {B(t)}}=\exp(-t\lambda_\theta)$. Now~\eqref{charGVG} follows from~\eqref{LaplaceGGCd} via conditioning on $T(t)$:
\[
E[\exp(\rmi\skal\theta {Y_t})]=E[e^{- T_t \lambda_\theta}]=\exp\big\{-t a\lambda_\theta-t\int_{(0,\infty)} \log[(x+\lambda_\theta)/x]\;\TTT(\rmd x)\big\}\,.\]
Here the right hand-side matches the formulae in~\eqref{charGVG}.\\[1mm]
(b)~\eqref{GVGlevdensI} is shown in~\cite{Gr07} (his Proposition~3.3), whereas ~\eqref{GVGlevdensII} holds as with $g_Y(s,r)=r^d\; \frac{\rmd \Pi_Y}{\rmd y}(rs)$
in \eqref{GVGlevdensII} ($r>0$, $s\in\mySS^d$) and any Borel set $A\subseteq \RR_*$ (see~\cite{Gr07}, his Equation~(4)) we have
\[\Pi_Y(A)=\int_A \frac{\rmd \Pi_Y}{\rmd y}(y)\,\rmd y=\int_{\mySS^d_E}\int_0^\infty \eins_{A}(rs)\,\frac{\rmd \Pi_Y}{\rmd y}(rs) r^{d-1}\,\rmd r\,\rmd s\,.\]
\halmos
\\[1mm]
\noindent{\em Proof of Theorem~\ref{theoGVGmulti}.}~(a)~We omit the proof as it is similar to the proof of Part~(a) of  Theorem~\ref{theoGVG}.\\[1mm]
(b)~We decompose $T$ into a superposition of independent subordinators $T=\sum_{J\subseteq\{1,\dots,d\}}T^J$ where
$T^\emptyset:=aI$,  and
\begin{equation}\label{TconeJ} T^J_t:=\sum_{0<s\le t}\eins_{C_J}(\Delta T_s)\,\Delta T_s \,,\qquad t\ge 0\,,\;\emptyset\neq J\subseteq\{1,\dots,d\}\,,\end{equation}
with $C_J$ as in~\eqref{coneJ}. Here $\Delta T(t)=T(t)-T(t-)$ for $t>0$. Also, $I:[0,\infty)\to[0,\infty))$ denotes the identity.
By Proposition~\ref{propdecomp}, we have $Y\eqd \sum_{J\subseteq\{1,\dots,d\}}Y^J$ where $(Y^J)$ is a family of independent L\'evy processes with $Y^\emptyset\eqd B\circ_{d}(aI)$ and $Y^J\eqd B\circ_{d}T^J\sim L^d(\gamma_J,0,\Pi_J)$ with $T^J\sim S^d(0,\Pi_{T}^J)$ for $J\neq\emptyset$. For $J\neq\emptyset$ we have $\TTT(C_J)=0\Leftrightarrow T^J\equiv 0\Rightarrow Y^J=0 \Leftrightarrow \Pi_J\equiv 0$.

To see~\eqref{GVGlevdensmultiI}, suppose $\det\Sigma>0$ and $J\neq \emptyset$ with $\TTT(C_J)>0$.
Note $T^J\sim S^d(0,\Pi_{T}^J)$ with, using its polar representation,
\[\rmd\Pi_{T}^J\,=\,\eins_{C_J\cap\Sdplus}(s)\;\eins_{(0,\infty)}(r)\;k(s,r)\,\alpha(\rmd s)\rmd r \big/r\,,\]
where $k(s,r)$ is the quantity in~\eqref{GGClevydensity}--\eqref{GGCkfunction}.
In view of Lemma~\ref{lemunifunimultisub},
\[\Pi_{J}(\rmd x)\,=\,\int_{C_J}P\big(\mu\diamond t+(\Sigma\diamond t)^{1/2} Z\;\in\; \rmd x\big)\;\Pi_{T}^J(\rmd t)\,,\]
where $Z$ is a $d$-dimensional standard normal vector.

As both $\det\Sigma>0$
and $\Pi_T^J(C_J)>0$, $\Pi_{Y}^J$ must be absolutely continuous with respect to $\ell_J$, admitting the following density, for $y\in V_J$,
\[\frac{\rmd \Pi_J}{\rmd \ell_J}(y)=\int_{C_J\cap \Sdplus}\int_{0}^\infty \int_0^\infty
\frac{\exp\big\{\!-\!r\tau\!-\!\frac 12\|y\!-\!r\mu\diamond s\|^2_{J,rs}\big\}}{r(2\pi r)^{\#J/2}\prod_{j\in J} \Sigma_{jj}^{1/2} s_j^{1/2}}
\rmd r\,\,\KKK(s,\rmd\tau)\,\alpha(\rmd s)\,.\]
Here we set $\|x\|^2_{J,c}:=\sum_{j\in J}x_j^2/(c\Sigma_{jj})$ for $c\in C_J, x\in\RR^d$. Expanding a square in the exponent yields \[\frac 12\|y\!-\!r\mu\diamond s\|^2_{J,rs}=\frac 1{2 r}\|y\|\|^2_{J,s}-\sum_{j\in J}\frac{y_j\mu_j}{\Sigma_{jj}}+\frac r2\|\mu\diamond s\|^2_{J,s}\,,\]
so as to evaluate the interior $\rmd r$-integral using the identity~\eqref{Bessel} for the modified Bessel function $K$ of the second kind. For $y\in\RR^d_*$ we get
\begin{eqnarray}\label{BesselGVGlevdensmulti}
\frac{\rmd \Pi_J}{\rmd \ell_J}(y)&=&2^{(2-\#J)/2}\;\pi^{-\#J/2}
\exp\big\{\sum_{j\in J}\mu_j y_j/\Sigma_{jj}\big\}\;\;\times
\\
&&{}\times\int_{C_J\cap \Sdplus}\int_{(0,\infty)}
\bigg[\{2\tau\!+\!\sum_{j\in J}s_j\mu_j^2/\Sigma_{jj}\}/\|y\|^2_{J,s}\bigg]^{\#J/4}\;\times\nonumber\\
&&{}\qquad\qquad
K_{\#J/2}\big(\big\{2\tau\!+\!\sum_{j\in J}s_j\mu_j^2/\Sigma_{jj}\big\}^{1/2}\,\|y\|_{J,s}\big)\,\frac{\KKK(s,\rmd\tau)\,\alpha(\rmd s)}
{\prod_{j\in J}\Sigma_{jj}^{1/2}s_j^{1/2}}
\,,\nonumber
\end{eqnarray}
where the RHS of~\eqref{BesselGVGlevdensmulti} matches~\eqref{GVGlevdensmultiI}. In~\eqref{GVGlevdensmultiI}, observe \[r^{\#J}(\rmd \Pi_Y^J/\rmd \ell_J)(rs)=g_J(s,r)\,,
\quad r>0,\;s\in \mySS^d_E\cap V_J\,,
\]
where the RHS matches~\eqref{GVGlevdensmultiII}. This completes the proof of Part~(b).\halmos

\subsection{Proofs for Subsection~\ref{subsecmoments}}~\label{subsecproofmoments}
\noindent{\it Proof of Proposition~\ref{propGGCmoments}.}~(a)~Let $0<q<1$. Pick $\varepsilon>0$ such that, for all $\tau>0$,
\begin{equation}\label{eqsmallmoments}\varepsilon^2 \tau^{-q}\; \eins_{\tau>1}\le \varepsilon\tau^{-q}\int_0^\tau r^{q-1}\,e^{-r}\;\rmd r\le  1\wedge \tau^{-q}\,.\end{equation}
By~\eqref{GGClevydensity}--\eqref{GGCkfunction}, we get from Fubini's theorem and a simple substitution that
\begin{eqnarray}
\int_{0<\|z\|\le 1} \|z\|^{q}\;\Pi_T(\rmd z)&=&\int_{\mySS^d_+}\int_{0<r\le 1}\int_{(0,\infty)} r^q\|s\|^q e^{-r\tau} \KKK(s,\rmd\tau)\frac {\rmd r}r\,\alpha(\rmd s) \nonumber\\
&=&\int_{[0,\infty)^d_*} \;\int_{0}^{1} r^{q-1}\,e^{-\|x\|r}\;\rmd r\;\TTT(\rmd x)\,.\nonumber\\
&=&\int_{[0,\infty)^d_*} \;\|x\|^{-q} \int_{0}^{\|x\|} r^{q-1}\,e^{-r}\;\rmd r\;\TTT(\rmd x)\,.\label{eqFubsmallmem}\end{eqnarray}
In view of ~\eqref{thorinmeasure} and~\eqref{eqsmallmoments}--\eqref{eqFubsmallmem}, $\int_{\|x\|>1} \TTT(\rmd x)/\|x\|^{q}$ is finite if and only if $\int_{0<\|z\|\le 1} \|z\|^q\;\Pi_T(\rmd z)$ is, completing the proof of~(a).\\[1mm]
(b)~Let $p,t>0$. Pick $\varepsilon>0$ such that, for all $\tau>0$,
\begin{equation}\label{eqlargemoments}\varepsilon^2 \tau^{-p}\; \eins_{0<\tau\le 1}\le \varepsilon\tau^{-p}\int_\tau^\infty r^{p-1}\,e^{-r}\;\rmd r\le \eins_{0<\tau\le 1}\tau^{-p}+ \eins_{\tau> 1} e^{-\tau}\,.\end{equation}
Using similar arguments as in the proof of Part (a), we get from~\eqref{GGClevydensity}--\eqref{GGCkfunction}, Fubini's theorem and a simple substitution that
\begin{equation}\label{eqFublargemem}\int_{\|z\|\ge 1} \|z\|^p\;\Pi_T(\rmd z)\,=\,\int_{[0,\infty)^d_*} \;\|x\|^{-p}\,\int_{\|x\|}^\infty r^{p-1}\,e^{-r}\;\rmd r\;\TTT(\rmd x) \,.\end{equation}
In view of~\eqref{thorinmeasure} and~\eqref{eqlargemoments}--\eqref{eqFublargemem}, we see that $\int_{\|z\|\ge 1} \|z\|^p\;\Pi_T(\rmd z)$ is finite if and only if $\int_{0<\|x\|\le 1} \TTT(\rmd x)/\|x\|^{p}$ is, completing the proof of (b).\halmos\\[1mm]
\noindent{\it Proof of Proposition~\ref{propVGGCmoments}.}~Let $k=d$. Let $B\sim BM^d(\mu,\Sigma)$ where $\Sigma$ is a diagonal matrix.
Let $Z=(Z_1\dots,Z_d)'\in\RR^d$ be a standard normal vector, that is a vector with independent standard normal
components. For $s\in[0,\infty)^d$ introduce $B^*(s):= (\Sigma \diamond s)^{1/2}Z$.
By self-similarity of $B-\mu I$ we can write $B(s)\eqd \mu\diamond s+B^*(s)=\mu\diamond s+(\Sigma\diamond s)^{1/2}Z$ for all (but fixed)  $s\in[0,\infty)^d$.

For $x=(x_k)_{1\le k\le d}\in\RR^d$ and $M=(m_{kl})_{1\le k,l\le d}\in\RR^{d\times d}$ the maximum norms are denoted by
$\|x\|_\infty=\max_k|x_k|$ and $\|M\|_\infty=\max_{k,l}|m_{kl}|$, respectively.
Let $\|\cdot\|_{op}$ be the operator norm of $\|\cdot\|$.
The equivalence of norms in finite dimensions applies to $\RR^d$ as well as $\RR^{d\times d}$: we find a common constant $C_\infty\in[1,\infty)$ such that
$\|\cdot\|_{\infty}\le C_\infty \|\cdot\|$ as well as $\|\cdot\|_{op}\le C_\infty \|\cdot\|_\infty$. Further, with $\|\cdot\|_E$ denoting the Euclidean norm on $\RR^d$ there exists  $C_E\in(0,\infty)$ such that $\|\cdot\|\le C_E\|\cdot\|_E$.\\[1mm]
(a)~Let $0\!<\!q\!<\!2$, and introduce a function $h:[0,\infty)^d_*\to [0,1]$ by
\[h(s):=E[\|B(s)\|^q\eins_{(0,1]}(\|B(s)\|)]\,,\qquad s\in[0,\infty)^d_*\,.\]
`\eqref{intTTTqhalffinite}$\Rightarrow$\eqref{intPIYqfinite}':~We get from the self-similarity of $B-\mu I$ that
\[h(s)\,\le\,E[\|B(s)\|^q]=E\big[\big\|\mu \diamond s+(\Sigma\diamond s)^{1/2}Z\big\|^q\big]\,,\quad s\in[0,\infty)^d\]
and, thus, with $C_1:=2^{q\vee 1}C_\infty^{2q}\;\big(\,\|\mu\|^q+ E[\|\Sigma^{1/2}Z\|^q]\big)$,
\begin{eqnarray*}
h(s)&\le&2^{q\vee 1}\;\big(\|\cdot\diamond s\|^q_{op}\,\|\mu\|^q+\|\cdot\diamond s^{1/2}\|_{op}^{q}\, E[\|\Sigma^{1/2}Z\|^q]\big)\nonumber\\
&\le&2^{q\vee 1}C^q_\infty\;\big(\|\cdot\diamond s\|^q_\infty\,\|\mu\|^q+\|\cdot\diamond s^{1/2}\|_{\infty}^{q}\, E[\|\Sigma^{1/2}Z\|^q]\big)\nonumber\\
&=&2^{q\vee 1}C^q_\infty\;\big(\|s\|^q_\infty\,\|\mu\|^q+\|s\|^{q/2}_\infty\, E[\|\Sigma^{1/2}Z\|^q]\big)\nonumber\\
&\le&C_1\big(\eins_{0<\|s\|\le 1}\|s\|^{q/2}+\eins_{\|s\|>1}\|s\|^q\big)\,,\quad s\in [0,\infty)^d_*\,.
\end{eqnarray*}
As $h$ is globally bounded by $1$, there thus exists $C_2\in (0,\infty)$ such that $h(s)\le C_2 (\|s\|^{q/2}\wedge  1)$ for $s\in[0,\infty)^d_*$. \eqref{intPIYqfinite} follows from this and the finiteness of $\int_{0<\|s\|\le 1}\|s\|^{q/2}\;\Pi_T(\rmd s)$. (Recall~$\Pi_Y(\rmd y)= P(B(s)\in \rmd y)\,\Pi_T(\rmd s)$ by Lemma~\ref{lemunifunimultisub}.) The proof is completed by Part~(a) of Proposition~\ref{propGGCmoments}.\\[1mm]
`\eqref{intPIYqfinite}$\Rightarrow$\eqref{intTTTqhalffinite}':~Assume $\det \Sigma\neq 0$. For $s\in[0,\infty)_*^d$ observe
$\|\Sigma s\|_\infty=\max_k \Sigma_{kk} s_k>0$ and $P(Z\in\RR^d_*)=P(\Sigma^{1/2}Z\in\RR^d_*)=P(B^*(s)\in\RR^d_*)=1$
such that, since $\|Z\|^2_E\sim \chi^2_d=\Gamma(d/2,1/2)$, for any $\rho\in\RR$,\begin{eqnarray}
\lefteqn{\hspace*{-2cm}P\big\{\exp\{\rho \|B^*(s)\|_E\}\,\big\|B^*(s)\big\|^q\,\eins_{0<\|B^*(s)\|<1})=0\big\}=P\{\|B^*(s)\|\ge 1\}}&&\nonumber\\
&\le&P\{\|Z\|^2_E\ge (C_E^2\|\Sigma s\|_\infty)^{-1}\}\quad<\quad1\,.\label{eqsmallmomentsVGGCI}\end{eqnarray}
For $\lambda:=-\Sigma^{-1}\mu$ and $s\in[0,\infty)^d_*$ we get from Girsanov's theorem that
\[B^*(s)-\mu\diamond s=B^*(s)+(\Sigma\diamond s)\lambda\sim\LLL\big(B^*(s)\big|\exp\big\{\!\skal\lambda{B^*(s)}-\frac 12\|\lambda\|^2_{\Sigma\diamond s}\big\}P\big)\,,\]
and, by self-similarity of $B-\mu I$,
\begin{eqnarray}
h(s)&=&E\big[\|\mu\diamond s +B^*(s)\|^q\,\eins_{0<\|\mu\diamond s +B^*(s)\|\le 1}\big]\nonumber\\
&=&\exp\big\{-\|\lambda\|^2_{\Sigma\diamond s}/2\big\}\; E\big[\exp\{\!-\skal\lambda{B^*(s)}\}\;\|B^*(s)\|^q\;\eins_{0<\|B^*(s)\|\le 1}\big]\nonumber\\
&\ge&\|s\|^{q/2}\exp\{-\|s\|_\infty\|\lambda\|^2_{\Sigma}/2\}\;h_0(s/\|s\|)\nonumber\\
&\ge&\|s\|^{q/2}\exp\{-C_\infty\|\lambda\|^2_{\Sigma}/2\}\;h_0(s/\|s\|)\,,\quad
\label{eqsmallmomentsVGGCII}
\end{eqnarray}
for $s\in[0,\infty)^d_*$ with $\|s\|<1$. (Note $\|s\|_\infty\le C_\infty \|s\|\le C_\infty$ for $\|s\|\le 1$.)
Here $h_0:\mySS^d_+\to [0,\infty)$ is defined by
\[h_0(s):=E\big[\exp\{-\|\lambda\|_E\,\|B^*(s)\|_E\}\;\|B^*(s)\|^q\;\eins_{0<\|B^*(s)\|<1}\big]\,.\]
As $P(\prod \Sigma_{kk}^{1/2}Z_k\neq0)=1$, $s\mapsto\eins_{(0,1)}(\|B^*(s)\|)$ is lower semicontinuous on $\mySS^d_+$, almost surely (indicator functions of open sets are lower semicontinuous). As a result, $h_0$ is itself lower semicontinuous by Fatou's lemma with a compact domain $\mySS^d_+$. In particular, $h_0$ attains its global minimum at some $s_0\in\mySS^d_+$.
Note $h_0(s_0)>0$ by~\eqref{eqsmallmomentsVGGCI}. To summarise, we get from~\eqref{eqsmallmomentsVGGCII} that $h(s)\ge C_3\|s\|^{q/2}\eins_{0<\|s\|<1}$ for $s\in[0,\infty)^d_*$
with $C_3:=h_0(s_0)\exp\{-C_\infty\|\lambda\|^2_\Sigma/2\}\in(0,\infty)$. The proof is completed by an application of Part~(a) of Proposition~\ref{propGGCmoments}, using similar arguments as
 in the proof of~'\eqref{intTTTqhalffinite}$\Rightarrow$\eqref{intPIYqfinite}'.\\[1mm]
(b)~Let $t,p>0$. Recall $Y\eqd B\circ_d T$ for independent $B$ and $T$.\\
\noindent`\eqref{intTTTphalffinite}$\Rightarrow$\eqref{intpmomentYfinite}':~Set $C_4:=2^{p\vee 1}C_\infty^{2p}$. As in the proof of Part~(a) note
\begin{eqnarray}
E[\|B(s)\|^p]&\le&C_4\;\big\{\|s\|^p\,\|\mu\|^p+\|s\|^{p/2}\, E[\|\Sigma^{1/2}Z\|^p]\big\}\label{momentVGGCI}\\
&\le&C_4\;\big\{E[\|\Sigma^{1/2}Z\|^p]+\|s\|^p(\|\mu\|^p+E[\|\Sigma^{1/2}Z\|^p])\big\}\label{momentVGGCII}
\end{eqnarray}
for $s\in[0,\infty)^d$. If $\mu=0$ then it follows from~\eqref{momentVGGCI} that $E[\|Y(t)\|^p]=E[E[\|B(T(t))\|^p|T(t)]]\le C_4 E[\|\Sigma^{1/2}Z\|^p]\; E[\|T(t)\|^{p/2}]$, and the LHS is finite provided $E[\|T(t)\|^{p/2}]$ is. Otherwise, if $\mu\neq 0$, we get from~\eqref{momentVGGCII} that
\begin{eqnarray*}
E[\|Y(t)\|^p]
&\le& C_4 \big\{E[\|\Sigma^{1/2}Z\|^p]+E[\|T(t)\|^p]\big(\|\mu\|^p\!+\!E[\|\Sigma^{1/2}Z\|^p]\big)\}\,,\end{eqnarray*}
and the LHS is finite provided $E[\|T(t)\|^{p}]$ is. In view of Part~(b) of Proposition~\ref{propGGCmoments},
this completes the proof of `\eqref{intTTTqhalffinite}$\Rightarrow$\eqref{intPIYqfinite}'.\\[1mm]
`\eqref{intpmomentYfinite}$\Rightarrow$\eqref{intTTTphalffinite}':~Suppose $\mu=0$. Define $g:\mySS^d_+\to[0,\infty)$ by $g(s):=E[\|B^*(s)\|^p\eins_{\|B^*(s)\|>1}]$.
Employing similar arguments as in the proof of Part~(a) we find
$s_0\in \mySS^d_+$ such that $\inf_{\mySS^d_+}g=\min_{\mySS^d_+}g=g(s_0)>0$ and, thus,
\[E[\|B(s)\|^p]=E[\|B^*(s)\|^p]=\|s\|^{p/2}E[\|B^*(s/\|s\|)\|^p]\ge g(s_0) \|s\|^{p/2}\,,\]
for $s\in[0,\infty)^d_*$. This extends to $E[\|B(s)\|^p]\ge g(s_0) \|s\|^{p/2}$ for $s\in[0,\infty)^d$, including the origin. In particular,
this implies the inequality $E[\|Y(s)\|^p]\ge g(s_0) E[\|T(t)\|^{p/2}]$ by conditioning on $T(t)$. In view of Part~(b) of Proposition~\ref{propGGCmoments}, this completes the proof for $\mu=0$.

Assume $\prod_{k}\mu_k\neq 0$. By the equivalence of norms, we have $\|\cdot\|\le C_5\|\cdot\|_\infty$ for some $C_5\in(0,\infty)$ such that, for $s\in[0,\infty)^d$,
\[\|s\|\le C_5\,\frac{\min_k|\mu_k|}{\min_k|\mu_k|}\,\|s\|_\infty\le
C_5\|1/\mu\|_\infty\,\|\mu\diamond s\|_\infty\le C_\infty C_5\|1/\mu\|_\infty\,\|\mu\diamond s\|,
\]
and, with $C_6:=(C_\infty C_5\|1/\mu\|_\infty)^p2^{1\vee p}$, using the self-similarity of $B-\mu I$,
\begin{eqnarray*}
\|s\|^p&\le& (C_\infty C_5\|1/\mu\|_\infty)^pE[\|\mu\diamond s +B^*(s)-B^*(s)\|^p]\\
&\le& C_6E[\|B(s)\|^p]+C_6E[\|B^*(s)\|^p]\le C_6E[\|B(s)\|^p]+C_7\|s\|^{p/2}
\end{eqnarray*}
where $C_7:=C_\infty^{2p}C_6E[\|\Sigma^{1/2} Z\|^p]$. Thus, we can find
$r_0\in(0,\infty)$ such that $C_6E[\|B(s)\|^p]\ge \|s\|^p-C_7\|s\|^{p/2}\ge \|s\|^p/2$ for $s\in [0,\infty)^d$ with $\|s\|>r_0$.
To summarise, we have $\|s\|^p\le 2C_8E[\|B(s)\|^p]+r_0^p$ for $s\in[0,\infty)^d$ and, thus, $E\|T(t)\|^p\le 2C_8E[\|Y(s)\|^p]+r_0^p$, by conditioning, completing the proof of~`\eqref{intpmomentYfinite}$\Rightarrow$\eqref{intTTTphalffinite}'.\\ (We omit the proof for $k=1$ being similar but simpler.)\halmos
\subsection{Proofs for Subsection~\ref{subsecEsscher}}~\label{subsecproofEsscher}
\noindent{\it Proof of Proposition~\ref{propexpoGGCVGGC}.}
(a)~Let $\CCC_\lambda:=\{0\}\cup ([0,\infty)_*^d\backslash\OOO_\lambda)$. It is straightforwardly checked that $\CCC_\lambda$ is closed under taking convex combinations.
For $x\in\CCC_\lambda$ we have $\|x\|^2\le \skal \lambda x\le C\|\lambda\|_E \|x\|$ and, thus,
$\|x\|\le C \|\lambda\|_E $
by the Cauchy-Schwarz inequality ($\|\cdot\|_E$ denotes the Euclidean norm and $C\in (0,\infty)$ is any constant with
$\|\cdot\|_E\le C \|\cdot\|$.) Thus, $\CCC_\lambda$ is a bounded subset of $\RR^d$. In particular, $\CCC_\lambda$ is a compact, as it is
also a closed subset of $\RR^d$.

Continuity of $\SSS_\lambda$ is obvious. For $x\in \OOO_\lambda$ we have $\|x\|^2-\skal\lambda x>0$ and thus $\SSS_\lambda(x)\in[0,\infty)^d_*$,  as desired.\\[1mm]
(b)~Let $\lambda\in\RR^d$.
We get from Fubini's theorem and~\eqref{GGClevydensity} that
\begin{eqnarray}\nonumber
\int_{\|x\|>1} e^{\skal \lambda x}\Pi_T(\rmd x)&=&
\int_{\Sdplus}\int_{(0,\infty)}\int_{1}^\infty e^{r(\skal\lambda s-\tau)}\frac{\rmd r}r\;\KKK(s,\rmd \tau)\alpha(\rmd s)\\
&=&\int_{[0,\infty)^d_*}\!\int_{1}^\infty \exp\Big\{\!\!-r\frac{\|x\|^2-\skal\lambda x}{\|x\|}\Big\}\frac{\rmd r}r\;\TTT(\rmd x)\,.\nonumber\\
\label{idexpomom0}
\end{eqnarray}
Consequently, if $\TTT([0,\infty)^d_*\backslash\OOO_\lambda)>0$ then $\lambda\notin\DDD_T$. For the remaining part, assume $\TTT([0,\infty)^d_*\backslash\OOO_\lambda)=0$, and choose $\varepsilon>0$ such that for all $\tau>0$
\begin{equation}\label{ineqexponmom}\varepsilon^2 \log^-(\tau)
\le\varepsilon\int_{\tau}^\infty e^{-r}\frac{\rmd r}r\le
\log^-(\tau)+e^{-\tau}\,.
\end{equation}
Note that
\begin{equation}
\int_{\OOO_\lambda}\exp\{(\skal\lambda x-\|x\|^2)/\|x\|\}\TTT(\rmd x)\le \sup_{s\in \mySS^d_+}e^{\skal\lambda s}\times
\int_{\OOO_\lambda}e^{-\|x\|}\,\TTT(\rmd x)\,.
\label{idexpomom}
\end{equation}
In~\eqref{idexpomom} the right hand-side is finite in view of~\eqref{thorinmeasure}.
The proof of Part~(a) is completed by combining \eqref{idexpomom0}, \eqref{ineqexponmom} and~
\eqref{idexpomom}.\\[1mm]
(c)~Suppose $\lambda\in \DDD_T$. (By Part (b) we have $\TTT([0,\infty)^d_*\backslash\OOO_\lambda)=0$.)
In view of Part~(a), $\SSS_\lambda:\OOO_\lambda\to [0,\infty)^d_*$
is Borel measurable. In particular, the image measure, denoted by $\TTT_\lambda=(\TTT\big|\OOO_\lambda)\circ\SSS_\lambda^{-1}$, of the restiction of $\TTT$ to $\OOO_\lambda$ under $\SSS_\lambda$ is a well-defined Borel measure on $\RR^d_*$

With $\SSS_\lambda$ as in
\eqref{defS}, note that there is a constant $C\in(1,\infty)$ such that, for $x\in\OOO_\lambda$ with $\|\SSS_\lambda(x)\|\ge 1$,
\begin{equation}\label{xoverSSS}
\frac{\|x\|}{\|\SSS_\lambda(x)\|}=1+\frac{\skal \lambda x}{\|x\|^2-\skal \lambda x}\le
1+\frac{|\skal \lambda x|}{\|x\|^2-\skal \lambda x}\le1+\frac{|\skal \lambda x|}{\|x\|}\le C\,.\end{equation}
and, thus,
by the transformation theorem,
\[\int_{[0,\infty)^d_*}(1\!+\!\log^-\!\|x\|)\wedge \frac 1{\|x\|}\,\TTT_\lambda(\rmd x)=\int_{\OOO_\lambda} (1\!+\!\log^-\|\SSS_\lambda(x)\|)\wedge\frac 1{\|\SSS_\lambda(x)\|}\,\TTT(\rmd x)\,,\]
\begin{equation}\label{xoverSSSII}
\le C\int_{\OOO_\lambda} (1\!+\!\log^-\|x\|)\wedge\frac 1{\|x\|}\,\TTT(\rmd x)+\int_{\OOO_\lambda}\log^-\|\SSS_\lambda(x)\|\,\TTT(\rmd x)\,.\end{equation}
(To show the inequality, split $\OOO_\lambda$ into $\{x\in\OOO_\lambda:\|\SSS_\lambda(x)\|<1\}\cup\{x\in\OOO_\lambda:\|\SSS_\lambda(x)\|\ge 1\}$ and recall  $C>1$ in \eqref{xoverSSS}.) In view of~\eqref{thorinmeasure} and~\eqref{eqGGCd2expmoments}, the RHS in\eqref{xoverSSSII} is finite, completing the proof.\\[1mm]
(d)~Let $\lambda\in\RR^d$, $t>0$. If $k=d$, then we have $T\eqd B\circ_d T$ for independent
$T$ and $B\sim BM^d(\mu,\Sigma)$. Conditioning on $T(t)$ yields
\begin{eqnarray}\nonumber
E\exp\skal{\lambda}{Y(t)}&=&E\exp\skal\lambda {B(T(t))}
=E\exp\{\skal{\mu\diamond T(t)}\lambda+\frac 12 \|\lambda\|^2_{\Sigma\diamond T(t)}\}\\
&=&E\exp{\skal{q_{\lambda,d}}{T(t)}}\,.\label{EqEexpT} \end{eqnarray}
Otherwise, if $k=1$ then $E\exp\skal\lambda {Y(t)}=E\exp\{q_{\lambda,1} T(t)\}$.
In either way, this completes the proof of Part~(d). \halmos\\[2mm]
\noindent{\em Proof of Theorem~\ref{theoEsscherVGGC}.}~Let $k=d$, $t>0$, $\lambda\in\DDD_Y$. Let $q:=q_{\lambda,d}\in\RR^d$ as in~\eqref{defqlambda}. As $\lambda\in\DDD_Y$, we must have $q\in\DDD_T$ by Part~(d) of Proposition~\ref{propexpoGGCVGGC}.

Let $a=0$. Adapting arguments from the proof of Theorem~25.17 of~\cite{s}, $e.g.$, we get from~\eqref{LaplaceGGCd} that, for $z\!\in\!\CC^d$ with $q-\Re z\in[0,\infty)^d$,
\begin{equation}\label{inproofLaplaceGGCd}
E\exp\skal{z}{T(t)}\,=\,\exp\Big\{-t\int_{[0,\infty)d_*} \log\frac{\|x\|^2-\skal{z}{x}}{\|x\|^2}\;\TTT(\rmd x)\Big\}\,.
\end{equation}
Let $\OOO_q$ as in \eqref{defOlambda}, but with $\lambda$ replaced by $q$.
Observe $\TTT([0,\infty)^d_*\backslash \OOO_q)=0$, the latter by Part~(b) of Proposition~\ref{propexpoGGCVGGC}.
Note that $\SSS_q(x)/\|\SSS_q(x)\|^2=x/(\|x\|^2-\skal{q}{x})$ for $x\in\OOO_q$. Set $\mu_\lambda=\mu+\Sigma\lambda$.

Since~\eqref{EqEexpT} extends as well, we get from~\eqref{inproofLaplaceGGCd} that,
still with $a=0$,
\begin{eqnarray*}
&&Ee^{\skal{\lambda+\rmi \theta}{Y(t)}}\big/E e^{\skal{\lambda}{Y(t)}}\\
&=&\exp
\Big\{
-t\int_{\OOO_q}
\log\frac{\|x\|^2-\skal{q}{x}-\rmi \skal{\mu_\lambda\diamond x}{\theta}+\frac 12\|\theta\|^2_{\Sigma\diamond x}} {\|x\|^2-\skal{q}{x}}\;\TTT(\rmd x)
\Big\}
\\
&=&\exp
\Big\{
-t\int_{\OOO_q}
\log\frac{\|\SSS_q(x)\|^2-\rmi \skal{\mu_\lambda\diamond \SSS_q(x)}{\theta}+\frac 12\|\theta\|^2_{\Sigma\diamond \SSS_q(x)}} {\|\SSS_q(x)\|^2}\;\TTT(\rmd x)
\Big\}\,.
\end{eqnarray*}
Next, apply the transformation theorem to $\TTT\big|\OOO_q$ and $\SSS_q:\OOO_q\to[0,\infty)^d_*$ to see that the RHS of the last display matches~\eqref{charGVGmulti}, but with $a,\mu,\TTT$ replaced by $0,\mu_\lambda,\TTT_q$, respectively, where $\TTT_q$
is the well-defined Thorin measure in Part~(c) of Proposition~\ref{propexpoGGCVGGC}, but with $\lambda$ replaced by $q$.

According to~\eqref{charGVGmulti}, if $a\neq 0$ it is possible to decompose $Y\eqd B+Y_0$ into independent $B,Y_0$ where $B\sim BM^d(\mu\diamond a,\Sigma\diamond a)$ and $Y_0\sim \VGGC^{d,d}(0,\mu,\Sigma,\TTT)$. Using the independence, the proof is completed for $k=d$ by noting that \[E\exp\skal{\lambda+\rmi\theta}{B(t)}/E\exp\skal{\lambda}{B(t)}
=\exp\big\{t\rmi \skal\theta{\mu_\lambda\diamond a}-\frac t2 \|\theta\|^2_{\Sigma\diamond a}\big\}\,,\;\theta\in\RR^d\,.\]
The proof of the remaining case, where $k=1$, is similar, but simpler. This completes the proof of the theorem.\halmos
\newpage
\subsection*{Acknowledgment}
We thank the Editor, the Associated Editor and two Referees for a close reading of the paper and many helpful suggestions.
Further, we take pleasure in thanking Christian Rau for his comments on an earlier version of this manuscript.

\end{document}